\documentclass[11pt,a4paper]{article}
\usepackage[T1]{fontenc}
\usepackage{longtable}

\usepackage[a4paper,top=2cm,bottom=2cm,left=3cm,right=3cm,marginparwidth=1.75cm]{geometry}
\usepackage{multirow} 
\usepackage{amsmath}
\usepackage{graphicx}
\usepackage[font=footnotesize]{caption}
\usepackage{hyperref}
\usepackage{algorithm}
\usepackage{algpseudocode}
\usepackage{dsfont}
\usepackage{booktabs}
\usepackage{enumitem}
\usepackage{subcaption}
\usepackage{float} 
\usepackage{amssymb}
\usepackage[normalem]{ulem}
\usepackage{needspace}
\usepackage{soul}
\usepackage{threeparttable}
\usepackage{tabularx} 
\usepackage{array} 
\usepackage{makecell} 

\usepackage[affil-it]{authblk}

\makeatletter
\renewcommand{\section}{\@startsection {section}{1}{\z@}%
              {24pt}{12pt} {\large\scshape\bfseries}}
\renewcommand{\subsection}{\@startsection {subsection}{2}{\z@}%
             {12pt}{12pt}  {\itshape\bfseries}}
\renewcommand{\subsubsection}{\@startsection {subsubsection}{3}{\z@}%
             {10pt}{10pt}  {\normalsize\itshape}}
\makeatother

\usepackage{apacite}
\usepackage{natbib}

\usepackage{xcolor}

\title{\bfseries \normalsize Delphos: 
A reinforcement learning framework for assisting discrete choice model specification}

\author[1]{Gabriel Nova*}
\author[2]{Stephane Hess}
\author[1]{Sander van Cranenburgh}

\affil[1]{CityAI Lab, Transport and Logistics Group, Delft University of Technology, The Netherlands}
\affil[2]{Choice Modelling Centre - Institute for Transport Studies, University of Leeds, England}

\date{\vspace{-5ex}}

\begin{document}
\maketitle

We introduce Delphos, a deep reinforcement learning framework for assisting the discrete choice model specification process. Delphos aims to support the modeller by providing automated, data-driven suggestions for model specifications, thereby reducing the effort required to develop and refine utility functions. Delphos conceptualises model specification as a sequential decision-making problem, inspired by the way human choice modellers iteratively construct models through a series of reasoned specification decisions. In this setting, an agent learns to specify competitive candidate model specifications by choosing a sequence of modelling actions, such as adding alternative specific constants, accommodating both generic and alternative-specific taste parameters, applying non-linear transformations to attributes, and including interactions with covariates. Each resulting candidate model is estimated and evaluated using a reward function defined by the modeller, which can reflect statistical model fit (e.g., log-likelihood) as well as behavioural expectations (e.g., parameter signs). Specifically, Delphos uses a Deep Q-Network to learn how individual specification decisions contribute to the eventual quality of the resulting model and, in turn, which sequences of modelling decisions tend to produce well-performing candidates. We evaluate Delphos on both simulated and empirical datasets using alternative reward functions. In simulated cases, learning curves, Q-value patterns, and performance metrics show that Delphos learns effective specification strategies while exploring only a small fraction of the feasible modelling space. We further apply the framework to two empirical datasets to benchmark and demonstrate its practical use. These experiments illustrate the ability of Delphos to generate competitive, behaviourally plausible models and highlight the potential of this adaptive, learning-based framework to assist the model specification process.\\


\vspace{30ex}
\noindent\rule{\linewidth}{0.4pt}  
E-mail address: g.nova@tudelft.nl (G. Nova).

\clearpage
\section{Introduction}
Discrete Choice Models (DCMs) are widely used to understand individuals’ decision-making behaviour, to forecast demand, and to evaluate policies under various scenarios \citep{hess2005mixed, guevara2006endogeneity,hess2024handbook, mariel2021environmental,liebe2023maximizing}. These models aim to mathematically represent observed choice processes, which requires analysts to specify a model and estimate its parameters based on well-established behavioural theories and using available data  \citep{nova2025understanding} . Extracting meaningful insights from such models then requires the careful selection of an appropriate specification, an iterative process that is often time-consuming due to the many modelling decisions involved, such as variable selection, functional forms, and interaction terms \cite{van2022choice}. The predominant approach to model specification remains hypothesis-driven, whereby researchers iteratively propose and test competing model specifications, typically balancing goodness-of-fit measures with parsimony and behavioural expectations \citep{train2009discrete, ortelli2021assisted}. 
In practice, this process can be both labour- and time-intensive, as the number of possible specifications grows combinatorially even with a relatively small set of modelling decisions, such as adding constants, accommodating non-linearities, and accounting for observed heterogeneity through covariates \citep{beeramoole2023extensive}. Consequently, navigating this large specification space relies heavily on researchers’ prior knowledge and judgment, potentially introducing biases and leaving relevant model specifications unexplored.\\

Recent studies explore the use of meta-heuristic algorithms to automate the specification process and assist analysts.  These approaches adopt a fundamentally candidate-centric perspective, framing model specification as a combinatorial optimisation that is solved through iterative search over candidate models \citep{ghorbani2025enhanced}. Among these approaches, \cite{paz2019specification} apply Simulated Annealing, which explores local neighbourhoods of specifications to improve goodness-of-fit.  \cite{ortelli2021assisted} consider a Variable Neighbourhood Search, applying operators to several neighbourhoods to iteratively enhance local model performance. \cite{beeramoole2023extensive} propose a bi-level constrained optimisation framework that integrates prior assumptions into the search process to align specification with theoretical expectations. \cite{haj5195530grammar} use Grammatical Evolution to find specifications that evolve through genetic operators guided by a novel domain-specific grammar. Similarly, \cite{ghorbani2025enhanced} integrate Grammatical Evolution at the upper level with Iterative Singular Value Decomposition at the lower level of a bi-level model to improve the predictive reliability of utility-based choice models. Across these studies, the common thread is a candidate-centric optimisation paradigm: new specifications are generated by applying search operators to the current best candidates, and progress is measured through improvements in objective functions such as log-likelihood or information criteria. Most approaches maintain a set of high-performing (or Pareto non-dominated) candidates and generate new models by modifying these candidates or exploring their local neighbourhoods. Collectively, this line of work demonstrates both the potential and the growing interest in metaheuristics as tools for supporting discrete choice model specification.\\

However, this candidate-centric perspective differs from how choice modellers typically refine specifications in practice. Model specification is inherently a dynamic and iterative learning process in which choice modellers learn from accumulated outcomes (e.g., repeated non-convergence, poor fit), adapt their modelling strategy, and adjust which refinements to prioritise or avoid. Even when metaheuristics exploit information from past candidates, they operate algorithmically on complete specifications rather than learning a reusable decision rule for selecting modelling refinements step by step. Consequently, they do not accumulate modelling knowledge about which decisions tend to be effective in particular modelling contexts. This can lead to inefficient exploration, and more importantly, limit transferability when the dataset, objective function, or behavioural constraints change, even within the same application domain. Therefore, model specification calls for a shift towards a learning-based approach capable of acting as an intelligent assistant by acquiring, accumulating, and reusing modelling knowledge. \\

This paper introduces Delphos, a novel Deep Reinforcement Learning (DRL) framework that treats model specification explicitly as a sequential decision-making problem, mimicking how human choice modellers iteratively construct models through a series of modelling decisions. Rather than applying search operators to generate candidates and evaluating them by their immediate results, Delphos learns a specification strategy conditioned on the current partial utility specification. Through repeated sequences of modelling decisions, the agent learns which actions are promising for a given specification and uses this knowledge to guide subsequent decisions. This approach draws on the success of DRL in vast combinatorial spaces such as board games \citep{silver2017mastering}, protein folding \citep{jumper2021highly}, and neural architecture search\citep{zoph2016neural}, where agents avoid testing every possible combination by learning to expand promising partial structures, a strategy Delphos applies directly to sequential modelling decisions to construct a utility specification.\\

Delphos formalises the specification process as a Markov Decision Process (MDP), where the current partial specification defines the state and feasible refinements define the actions. When the agent submits a specification for estimation, a custom modelling environment estimates candidate models using Apollo functions \cite{hess2019apollo} and evaluates the resulting model according to a reward function defined by the modeller. Although transitions between specifications are deterministic, the quality of individual modelling decisions is not known in advance and can only be assessed once the resulting model has been estimated. To translate this sparse estimation feedback into a reusable, state-conditioned modelling decision rule, a Deep Q-Network (DQN) is trained to approximate the expected discounted return of each feasible action. Through repeated specification sequences, Delphos accumulates experience from both successful and unsuccessful modelling decisions and uses this knowledge to guide subsequent specification decisions. Although Delphos requires repeated model estimations within a predefined modelling space, the resulting specification strategy captures knowledge that can be reused throughout the search process. \\

The remainder of the paper is organised as follows. Section~\ref{sec: research framework} introduces the main reinforcement learning components of our framework (Subsection~\ref{sec: RL}) and formulates model specification as a Markov Decision Process (Subsection~\ref{sec: RL + DCM framework}), defining the state space, action space, agent architecture, modelling environment, and reward function. Section~\ref{sec: monte carlo exp} describes the experimental design, including simulated and empirical case studies, while Section~\ref{sec: results} presents the results. Finally, Section~\ref{sec: conclusion} discusses the main insights, methodological implications, limitations, and potential extensions.

\section{Research framework}\label{sec: research framework}
To employ reinforcement learning in the choice model specification problem, this section revisits the RL paradigm as the foundation for our framework. Section~\ref{sec: RL} provides an overview of the core components of RL and their role in defining the agent-environment interaction. Then, Section~\ref{sec: RL + DCM framework} introduces our proposed RL framework, in which we formulate the model specification as a Markov Decision Process that enables the agent–environment interaction. 

\subsection{Revisiting reinforcement learning paradigm}\label{sec: RL}

Reinforcement learning is a machine learning paradigm that focuses on sequential decision-making problems, wherein an agent interacts with an environment to learn which actions to take through trial-and-error to maximise accumulated long-term rewards \citep{bertsekas2019reinforcement}. Unlike supervised learning, which relies on fixed datasets to train models on predefined input–output mappings, RL trains agents to adapt their behaviour in uncertain environments based on data gained through agent–environment interaction. Thus, this process is typically formalised as a Markov Decision Process \citep{bellman1957markovian}, where the agent observes the current state of the environment, selects action(s), transitions to a new state, and receives rewards. This feedback guides the agent in improving its policy ($\pi_\theta$), which maps states to a probability distribution over actions and governs the agent's behaviour\citep{burnetas1997optimal, littman1994markov}.\\

\begin{figure}[ht]
    \centering
    \IfFileExists{RL_paradigm.png}{%
        \includegraphics[width=0.6\linewidth]{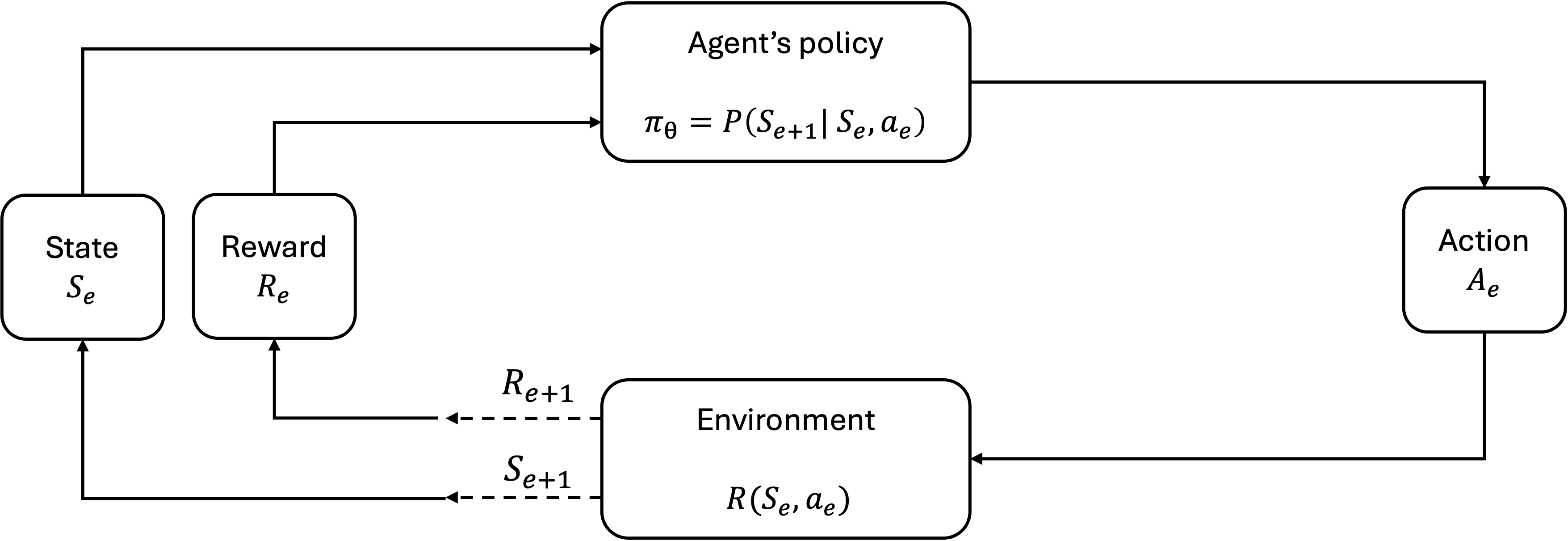}%
    }{%
        \IfFileExists{img/RL_paradigm.png}{%
            \includegraphics[width=0.6\linewidth]{img/RL_paradigm.png}%
        }{%
            \PackageWarningNoLine{Delphos}{Missing figure: RL_paradigm.png}%
            \fbox{\parbox[c][3cm][c]{0.6\linewidth}{\centering
                Missing figure: \texttt{RL\_paradigm.png}}}%
        }%
    }
    \caption{Classical RL framework. Adaptation from \cite{sutton2018reinforcement}}
    \label{fig: RL_paradigm}
\end{figure}

An MDP framework is defined as a tuple $(\mathbf{S},  \mathbf{A},  \mathbf{P}, \mathbf{R}, \mathbf{\gamma})$, where the state space ($\mathbf{S}$) contains all possible conditions of the environment; the action space ($\mathbf{A}$) defines the set of actions or operators that the agent can apply to transition from the current state to a new one; the transition probability  ($\pi_{\theta}(s_{e}, a_e) = P(s_{e+1} \mid s_e, a_e)$) models how the environment evolves from one state $s_e$ to another $s_{e+1}$ based on the agent’s action $a_e$; the reward function $R(s_e, a_e)$ reflects the quality of an action $a_e$ taken in a given state $s_e$; and the discount factor ($\gamma$) balances future cumulative rewards relative to intermediate ones. The agent's objective is then to learn an optimal policy $(\pi_\theta^*)$, parametrised by $\theta$, that maximises the expected cumulative reward $(J(\pi_\theta))$. This policy is a mapping function that determines which actions should be taken at each state based on a probability distribution over the feasible action space. To achieve optimal decision-making, the agent balances exploration (sampling new actions to improve its knowledge of the environment) and exploitation (choosing actions that have previously yielded high rewards). Therefore, at each interaction, namely episode $e$, the agent observes the current state $s_e$, selects an action $a_e$, receives a reward $r_e$, transitions to a new state $s_{e+1}$, and updates the policy $\pi_\theta$ accordingly, as shown in Figure~\ref{fig: RL_paradigm}.

Traditional approaches have relied on value-based methods, such as Q-learning \citep{watkins1992q}, and policy-based methods, such as Policy Gradient \citep{sutton1999policy}, for learning the optimal policy \cite{graesser2019foundations}.  Value-based methods estimate an optimal value function ($V(s)$), which represents the expected cumulative reward that an agent can achieve from a given state by following the optimal policy. Instead of directly learning the policy itself, these methods derive it by selecting actions that maximise the state–action value function $(Q(s, a))$  \citep{wong2023deep}, which evaluates the long-term expected returns of choosing action $a_e$ in state $s_e$:

\begin{equation}
     V^{*} (s) =  \max_{\pi} V^{\pi}(s)  \quad \Rightarrow \quad \pi (s )  = \arg\max_{a \in \mathbf{A}} Q^*(s, a)
\end{equation}

 In contrast, policy-based methods directly parametrise the policy function  $\pi_\theta$ and optimise it by adjusting the parameters $\theta$ to maximise the expected cumulative rewards \citep{sutton2018reinforcement, schulman2017proximal, schulman2015high}. Unlike value-based methods, these methods adjust their policy using gradient ascent, where the parameters are iteratively updated in a direction that increases the likelihood of selecting high-value actions \citep{sutton1999between}. The policy gradient theorem \citep{sutton1999policy} formalises this update rule, as shown in Eq.~(\ref{Eq: policy gradient}), where $\nabla_\theta \log \pi_\theta(a \mid s)$ indicates the direction in which to adjust the policy to increase the likelihood of selecting action $a$ at state $s$, while the action–value function $Q(s,a;\theta)$ quantifies the expected return of that action under the current policy, in other words, it reflects how valuable the action is \citep{schulman2017proximal}.

\begin{equation}
\nabla_\theta J(\theta) = \mathbb{E} \left[ \nabla_\theta \log \pi_\theta(a \mid s) \cdot Q(s, a; \theta) \right] \label{Eq: policy gradient}
\end{equation}

Traditional RL methods are computationally infeasible in environments with dynamically or high-dimensional evolving state or action spaces, as these methods rely on maintaining an explicit Q-value table that stores expected rewards for all possible state-action pairs \citep{plaat2022deep}. To address these limitations, Deep Reinforcement Learning was introduced, which uses deep neural networks to approximate value functions or policies \citep{mnih2016asynchronous,wang2016dueling}. \\

The pioneering Deep Q-learning algorithm replaced the traditional Q-value function with a neural network, integrating an experience replay buffer and a target network to stabilise training \citep{mnih2015human}. This architecture first demonstrated the power of deep RL by outperforming human performance in Atari games. It subsequent enabled AI milestones such as AlphaGo defeating the world champion \citep{silver2017mastering} and AlphaFold compressing decades of empirical protein discovery into months\citep{jumper2021highly}. Specifically, the agent approximates state-action values $Q(s,a;\theta)$, which estimates the expected return ($G_t$) of taking an action $a$ at state $s$ under a policy $\pi_\theta=P(a|s)$, as shown in Eq.~(\ref{Eq: Q_value}). This approximation is made using a fully connected neural network, where the parameters $\theta$ are learned through training. The network encodes abstract representations of states and uses them to calculate Q-values for all possible actions. At each step, the agent selects the action with the highest Q-value to transition to a new state by interacting with the environment.

\begin{equation}\label{Eq: Q_value}
Q(s,a;\theta) = \mathbb{E} \left[  G_t  \mid  S_t = s,  A_t = a, \pi=\pi_\theta \right]\quad \text{where} \quad G_t = \sum_{k=0}^{\infty} \gamma^k r_{t+k+1}
\end{equation}

To improve and stabilise the learning, the agent uses both an experience replay buffer and a target neural network. This buffer stores past interactions as tuples $(s,a,r,s^{\prime})$, capturing current state, action taken, reward received, and subsequent state. During training, the agent samples from this buffer randomly rather than relying on consecutive experiences to update the policy weights, which breaks the temporal correlation that can destabilise action–value updates \citep{lin1992self, zhang2017deeper}. In addition, the target network is used to stabilise Q-value estimation and reduce large oscillations during weight updates \citep{mnih2013playing, ly2024elastic}. Instead of using a single network to estimate the current and future Q-values, the target network maintains a separate copy of the policy weights, which are periodically updated. As shown in Eq.~(\ref{Eq: Q_target}), while the policy network $Q(s,a;\theta)$ is updated at every training step (episode), the target network $Q(s',a';\theta^-)$ keeps its parameters fixed for a number of training steps before synchronising with current policy weights \citep{tsitsiklis1996analysis, mnih2015human}.

\begin{equation}\label{Eq: Q_target}
     y_{target}  =  r + \gamma \max_{a'} Q(s', a'; \theta^-)
\end{equation}

To train the agent and learn an optimal policy, the policy weights are iteratively updated using past interactions to reduce the distance between the predicted Q-values and their target values, following a temporal difference learning approach \citep{sutton2018reinforcement}. The objective is then to minimise the loss function (Eq.~(\ref{Eq:Q_loss}), where the gradient of the loss function  $\nabla_\theta \mathcal{L}(\theta)$ guides the update direction, and the learning rate $\alpha$ controls the step size. 

\begin{equation}\label{Eq:Q_loss}
\mathcal{L}(\theta) =  \mathbb{E}_{(s,a,r,s') \sim \mathcal{D}} \left[ \left(  y_{target}  - Q(s, a; \theta) \right)^2 \right]
\end{equation}

\begin{equation}\label{Eq: update}
\theta \leftarrow \theta - \alpha \nabla_\theta \mathcal{L}(\theta)
\end{equation}

The DQN agent is enabled to solve this type of learning task not only by sampling past interactions to update its policy weights, but also by relying heavily on an exploration strategy \citep{mnih2015human}. This strategy controls how the agent balances the trade-off between exploiting areas of the state space known to yield high rewards and exploring new regions to avoid becoming trapped in local optima. In our framework, the agent follows a so-called $\epsilon$-greedy policy, which selects a random action with probability $\epsilon$ and the action with the highest predicted Q-value with probability $1 - \epsilon$, which decreases linearly over episodes. The $\epsilon$-greedy exploration strategy is simple yet effective, as it promotes early exploration during training and gradually shifts towards more informed decisions as the agent accumulates experiences and updates its policy \citep{dabney2020temporally}.

\subsection{Automating discrete choice model specification using RL methods}\label{sec: RL + DCM framework}
\begin{figure}[ht]
    \centering
    \includegraphics[width=\linewidth]{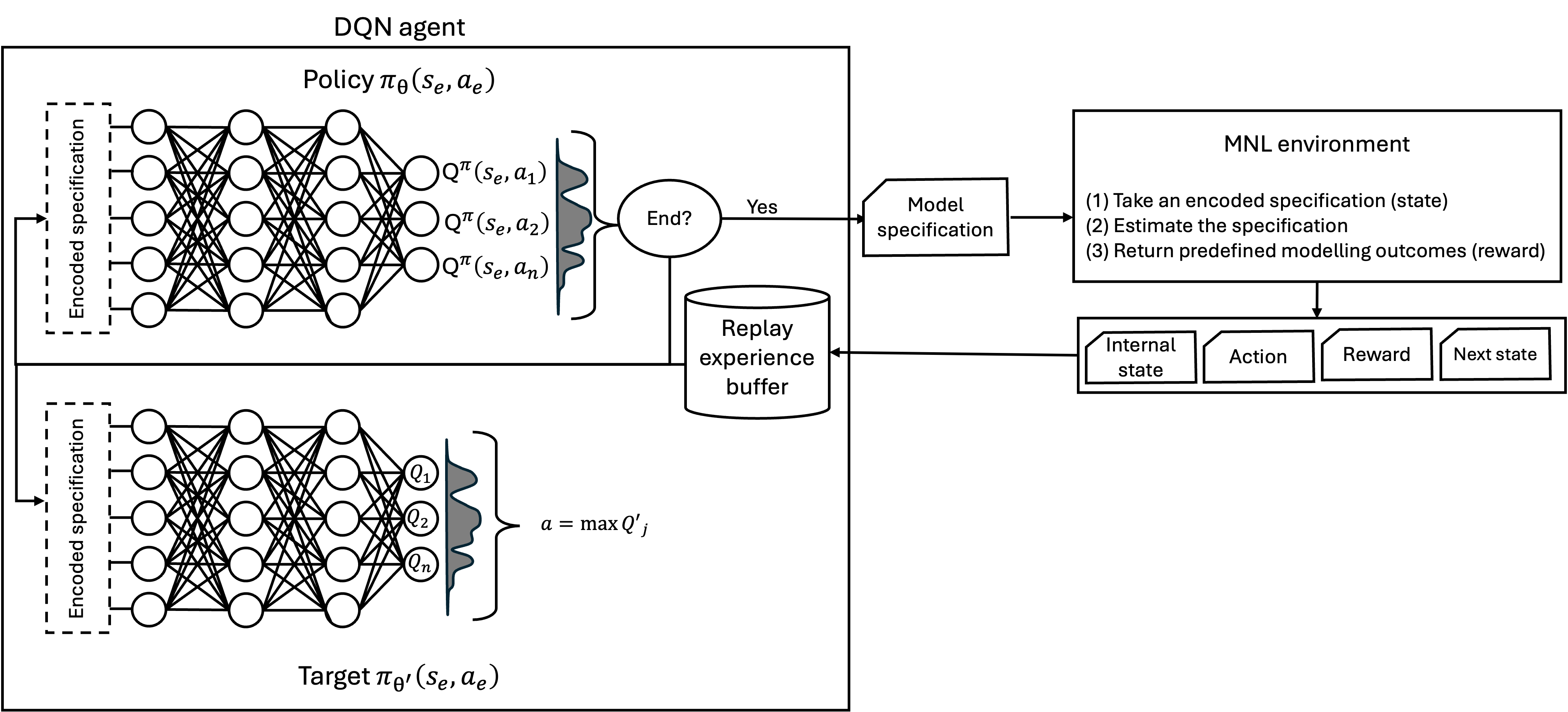}
    \caption{Framework for assisting discrete choice model specification. While Delphos encodes the current specification, selects a modelling action, and submits the model for estimation, the environment returns modelling outcomes as a reward signal. Transitions are stored in a replay buffer and used to update the policy and target networks. }
    \label{fig: DQN_DCM_framework}
\end{figure}

The proposed reinforcement learning framework formalises the agent–environment interaction to automate the search process for discrete choice models. This interaction is modelled as a Markov Decision Process, in which Delphos uses a Deep Q-Network to sequentially take actions that propose modifications to a model specification. The environment then estimates the resulting candidate and returns feedback based on the modelling outcomes, as depicted in Figure~\ref{fig: DQN_DCM_framework}.\\

Delphos autonomously builds model specifications by selecting attributes, applying transformations, incorporating interactions with covariates, or deciding when to estimate the model. Once a candidate specification is submitted, the environment estimates it, stores the corresponding modelling outcomes, and returns a reward signal that guides the agent’s learning process. Through multiple episodes, the agent learns to specify models by balancing exploration of new candidates with exploitation of well-performing ones to maximise cumulative rewards. 

\subsubsection{Problem formulation}
We translate the model specification task as a sequential decision-making problem within an MDP framework, where the agent learns to propose well-performing models —for instance, those that achieve high log-likelihood values and satisfy behavioural expectations such as expected parameter signs— by interacting with an estimation environment. At each episode, the agent takes a sequence of actions to build a specification and explore the modelling space, while the environment estimates the proposed candidate and returns a reward signal. \\

The environment contains a choice dataset that the agent can interact with. This dataset consists of $N$ observations representing decision-makers who face a choice situation among $J$ alternatives. Each alternative is described by $K$ attributes, $X = \{x_{nj1},\ldots, x_{njK}\}$, such as travel time, travel cost, or comfort. Additionally, it may include individual-specific covariates, $C = \{\text{cov}_{n1}, \ldots, \text{cov}_{nC}\}$, capturing socio-demographics or contextual information such as income, gender, or household size. The observed outcome $y_n$ indicates the alternative chosen by individual $n$.\\

While the current implementation of Delphos focuses on utility-based discrete choice models—where individuals are assumed to select the alternative with the highest utility—it can be readily extended to accommodate alternative decision rules, such as regret minimisation \citep{chorus2010new, van2015new}, satisficing \citep{swait2001non}, or rule-based approaches \citep{hess2013mixed}. In utility-based models, individual $n$'s utility $U_{ni}$ for alternative $i$ is determined by a deterministic component ($V_{ni}$), derived from observed attributes, covariates, and their associated parameters, and an error term ($\varepsilon_{ni}$) that captures unobserved factors, as shown in Eq.~(\ref{Eq: utility}).

\begin{equation} \label{Eq: utility}
    U_{ni} = V_{ni} + \varepsilon_{ni},
\end{equation}

This agnostic formulation allows our RL framework to be used with standard choice models, such as the Multinomial Logit (MNL), Generalised Extreme Value models (GEV)  \citep{mcfadden1977modelling}, Mixed Multinomial Logit (MXL) \citep{mcfadden2000mixed}, and Latent Class (LC) models \citep{walker2007latent}, among others. The present implementation of Delphos is restricted to the MNL. This restriction reflects the proof-of-concept nature of this study: Delphos is designed to demonstrate the potential of a DRL-based approach to assisting discrete choice model specification, rather than to provide a comprehensive model specification assistant. Focusing on the MNL allows us to study this potential in a controlled setting while retaining many of the core utility-specification decisions encountered in more advanced choice models, including the inclusion of attributes, non-linear transformations, alternative-specific parameters, and interactions with covariates. Moreover, many utility-specification insights identified within an MNL can directly inform the specification of richer model classes. For example, an interaction between an attribute and a socio-demographic covariate identified as relevant in an MNL specification may also be relevant when the same utility function is embedded within an MXL model that additionally accommodates random taste heterogeneity. Extending Delphos beyond the MNL would retain these utility-specification decisions while introducing additional decisions specific to the model family, such as which taste parameters are random and their distributions in an MXL, or the nesting and class structures in NL and LC models, respectively. Such decisions could be incorporated as additional actions and state information within the same sequential MDP structure, with action masks used to prevent invalid model specifications. Nevertheless, extending Delphos to these model families is not merely an implementation exercise, as each requires appropriate identification constraints and adaptations to the estimation environment, thereby increasing the computational cost of evaluating candidate models. These extensions are therefore left for future research. For the present MNL implementation, the modeller defines the modelling space through a set of modelling space parameters:

\begin{enumerate}
    \item \textbf{Variables:} A list of alternative-specific constants and attribute names that may enter into the utility specification, which is denoted as $\left [ASC, att_1, att_2, \ldots, att_K \right ]$. 

    \item \textbf{Transformations:} A predefined set of functional forms that can be applied to attributes to capture potential non-linearities. This list is denoted as $\left [f_1, f_2, \ldots, f_T \right ]$, where $T$ is the number of transformations. Among them are:

    \begin{itemize}
        \item Linear: $f_1(x) = x$
        \item Logarithmic: $f_2(x) = \log(1+x)$
        \item Box-Cox: $f_3(x) = (x^\lambda - 1)/\lambda$, where $\lambda$ is an extra estimated parameter.
    \end{itemize}

    \item \textbf{Taste parameters:} A list that indicates whether the attributes' parameters are considered either generic, alternative-specific taste parameters, or both. While a generic taste parameter is shared across all alternatives, an alternative-specific taste parameter varies across alternatives. This list is denoted as $\left [\text{generic}, \text{specific} \right ]$.

    \item \textbf{Covariates:} A list of variables that can be used to define interactions with alternative-specific constants or attributes to capture observed heterogeneity. This list is denoted as $\left [cov_1, cov_2, \ldots, cov_C \right ]$. 
\end{enumerate}

For instance, a modelling space that enables non-linear transformations to attributes, generic and alternative-specific taste parameters, and allows interactions with covariates, may be defined as:

\begin{align}
    \text{modelling space} =
    \begin{cases}
        \text{variables} &= [\text{asc, $att_1$, ...,$att_K$}]\\
        \text{transformations} &= [\text{linear}, \text{logarithm}, ..., f_{t} ]\\
        \text{taste} &= [\text{generic, specific}]\\
        \text{covariates} &= [\text{$cov_1$, ..., $cov_C$}]\\
    \end{cases}
\end{align}

Given this problem and the defined modelling space, we now formalise the elements that govern the decision-making process and the agent itself for specifying discrete choice models. These elements include the state space (Subsection~\ref{sec: state space}), the action space (Subsection~\ref{sec: action space}), the reward function (Subsection~\ref{sec: reward function}), the environment (Subsection~\ref{sec: environment}), and the agent (Subsection~\ref{sec: agent}).

\subsubsection{The state space}\label{sec: state space}
The state space contains all model specifications that can be built from the predefined modelling space, whereas an individual state represents the current model specification at an intermediate step of the agent’s decision process. We encode states as a fixed-length list of modelling terms (tuples), each describing a selected variable and its current transformation, taste parameter type, and covariate interaction. This concise yet flexible representation facilitates the application of any action in any order, thereby enabling the agent to propose any feasible model within the modelling space. Formally, a state $S_e^l$ represents the model specification resulting from the actions taken up to intermediate step $l$ within episode $e$ and is defined as:

\begin{equation}\label{Eq: state_1}
    S_e^l  =  \bigg \{ (k_1,t_1,g_1,i_1),\ldots,(k_q,t_q,g_q,i_q) \bigg \}_{q=1}^{K+1}.
\end{equation}

where $e$ denotes the model specification attempt and $l$ denotes the intermediate step within that episode. The state contains one tuple for each of the $K$ attributes and the alternative-specific constant, giving it a fixed length of $K+1$. For the $q$th modelling term, $k_q$ denotes the variable index, while $t_q$ identifies its transformation: $t_q=1$ denotes a linear specification, $t_q=2$ a logarithmic transformation, and $t_q=3$ a Box--Cox transformation. The value $g_q$ identifies the taste parameter type, with $g_q=1$ denoting a generic parameter and $g_q=2$ an alternative-specific parameter. Finally, $i_q$ identifies the covariate interaction: $i_q=0$ denotes no interaction, whereas $i_q\in\{1,\ldots,C\}$ identifies the selected covariate. When variable $k_q$ is excluded from the current specification, its configuration is encoded as $(k_q,t_q,g_q,i_q)=(k_q,0,0,0)$; while any other configuration describes how it enters the current model. Therefore, the modelling terms defining the modelling space remain fixed throughout training, while the configuration of each modelling term included in the current state evolves as the agent takes actions.

\subsubsection{The action space}\label{sec: action space}
The action space defines all feasible operations that the agent can perform to modify the current state (i.e., the current model specification), which is defined before training and remains fixed. These include  \textit{adding} new variables, \textit{changing} existing specification components, or  \textit{terminating} the specification process and submitting the model for estimation. Thus, at each intermediate step $l$ during episode $e$, the agent selects an action to modify any component of the list of modelling terms that define the state $S_e^{l-1}$ and to transition to $S_e^{l}$.\\

To prevent redundant or infeasible actions, Delphos constructs a state-dependent mask over the fixed action space. For instance, an \textit{add} action is invalid when an attribute is already included, while a \textit{change} action is invalid when the attribute is excluded, or the proposed modification is identical to its current configuration. The set of valid actions is thus dynamically masked based on the current intermediate state (i.e., we assign extremely negative Q-values to these actions, making them unlikely to be selected). This masking mechanism ensures that the agent takes meaningful actions (e.g., changing a component of an already added attribute instead of adding it again), which improves learning efficiency but makes the action space dynamic \citep{huang2020closer}. Formally, the agent can perform any of the following actions:

\begin{itemize}
    \item  \textit{Add} a new encoded variable  $k_q$ to expand the model $(k_q,1,1,0)$. While alternative-specific constants are included without interactions and one of them is fixed, the attributes are linearly incorporated with the generic taste parameter across alternatives and without interactions.

    \begin{equation}
    \text{Add}(S_e^{l}, k_q^l, t_q^l, g_q^l, i_q^l) \rightarrow S_e^{l+1} = S_e^l \cup \{(k_q^l,1,1,0)\} \,\,\, \forall k_q \notin S_e^l 
    \end{equation}

    \item  \textit{Change} a component of an existing variable in the model. This action allows modifications to any element that defines the modelling term tuple, such as its transformation, taste parameter specification, or interaction with a covariate. For example, changing the transformation of variable $k_q$ from $t_q$ to $\bar{t}_q$, while keeping the remaining components unchanged, is defined as    

    \begin{equation} \label{Eq: action_change}
    \begin{aligned}
    \operatorname{Change}(S_e^l,k_q,\bar{t}_q)
    \rightarrow
    S_e^{l+1} &= \left(S_e^l\setminus{(k_q,t_q,g_q,i_q)}\right)\cup
    {(k_q,\bar{t}_q,g_q,i_q)}
    \end{aligned}
    \end{equation}

    \item  \textit{Terminate} the model specification process. It thus leaves the final state unchanged  and submits the specification $S_e^l$ for its estimation.
 \end{itemize}

As with the state space, future work could expand the range of actions to include more operations, such as defining interactions between attributes themselves, considering categorical treatment, or adjusting parameter distributions, among others. These actions can then be incorporated by expanding the fixed action space and its masking rules.

\subsubsection{The reward function}\label{sec: reward function}
The reward function translates the modeller's goals into a learning signal that guides the agent's specification decisions. In Delphos, this signal provides feedback on the quality of the model resulting from a sequence of modelling actions and determines model evaluation criteria are prioritised. Depending on the modeller's objective, these priorities may emphasise goodness-of-fit, parsimony, or behavioural plausibility. Moreover, modellers often rely on multiple modelling outcomes to guide the specification process. For example, a modeller may aim to maximise both in-sample and out-of-sample log-likelihood to mitigate overfitting or balance goodness-of-fit against parsimony \citep{nova2025understanding}. The reward formulation therefore represents a deliberate design choice that shapes the policy learned by the agent \citep{dewey2014reinforcement}. \\

To reflect different modelling objectives, Delphos allows the modeller to define the episodic reward ($R_e$) by combining selected model evaluation criteria. These may include, but are not limited to, log-likelihood ($LL$), McFadden's pseudo-$R^2$ metrics  ($\rho^2$ and $\bar{\rho}^2$), Akaike Information Criterion ($AIC$), Bayesian Information Criterion ($BIC$), estimates, standard errors, number of estimated parameters, convergence information, and estimation time. In general, considering a set of selected modelling outcomes $M$ and  their corresponding weights ($\omega_m$), the sparse terminal reward associated with the specification obtained at the end of episode $e$ is defined as

\begin{equation}\label{eq: reward}
    R_e  =   \left [ \sum_{m \in M} \omega_m\cdot\tilde{M}_{m,e} \right ] \cdot \mathds{1}_{\text{converged}} \cdot \mathds{1}_{\text{behavioural}}
\end{equation}

where $\widetilde{M}{m,e}$ denotes the scaled value of modelling outcome $m$. The indicator $\mathds{1}{\text{converged}}$ equals one when the candidate model is successfully estimated, while convergence issues are treated as undesirable modelling outcomes to the agent. However, such failures are also informative and remain part of the learning process, especially in early training episodes, as they help the agent avoid poor modelling decisions \citep{ross2010efficient}. Similarly, Delphos can incorporate behavioural constraints to align its learning with behavioural and theoretical expectations. For example, a specification can receive zero reward when an estimated travel time coefficient violates an expected negative sensitivity. These constraints are implemented through $\mathds{1}_{\text{behavioural expectations}}$, which sets the reward to zero whenever a behavioural expectation is violated based on the estimation results.\\

Similar to a human choice modeller's workflow, Delphos sequentially modifies a model specification and estimates the resulting candidate once the terminate action is selected. Because these outcomes can only be evaluated after a complete specification has been estimated, the agent receives delayed feedback and must associate earlier decisions with their eventual effect on model performance. To address this temporal credit assignment, the final episodic reward is propagated across the sequence of $L$  modelling actions \citep{kaelbling1996reinforcement,sutton2018reinforcement}. For an action taken at step $\ell\in{1,\ldots,L}$, the corresponding discounted return $G_e^l$ assigned to that state-action pair is

\begin{equation}\label{eq: discounted_return}
    G_e^l = \gamma^{L - l} \cdot R_e
\end{equation}

where $\gamma\in[0,1]$ is the discount factor. Consequently, actions closer to termination receive greater credit, while earlier actions still contribute meaningfully to the final results when intermediate rewards are not directly available \citep{zhong2018practical}. This not only allows the agent to learn how individual decisions within a sequence of modelling actions contribute to modelling performance, but also to follow a shorter sequence of actions.\\

To ensure that modelling outcomes with different numerical scales can be combined within the reward, Delphos scales each metric using theoretically meaningful bounds known before training. This ensures that the same specification receives the same reward regardless of when it is encountered, providing a stable learning signal throughout training \citep{mnih2015human,van2016deep}. Delphos uses the null log-likelihood ($LL_0$) as the baseline for model performance\footnote{Although a model may perform worse than the null, such solutions would not be considered valid maximum-likelihood estimates \citep{train2009discrete, mokhtarian2016discrete}.}, while the number of observations ($N$) is used to express each metric on a per-observation level, as shown in Table~\ref{tab:reward_scaling}.\\

\begin{table}[ht]
\centering
\caption{Scaling of the modelling outcomes used in the reward function.}
\label{tab:reward_scaling}

\small
\renewcommand{\arraystretch}{1.4}

\begin{tabularx}{\textwidth}{XXp{3cm}}
\toprule
Modelling outcome & Formulation & Bound \\
\midrule
Log-likelihood & $\displaystyle \max\left(0,\frac{LL-LL_0}{N}\right) $ & $\displaystyle [0,\log J]$ \\[6pt]
AIC & $\displaystyle \max\left(0, \frac{LL-LL_0}{N}-\frac{K}{N} \right)$ & $\displaystyle \left[ 0, \log J-\frac{K}{N} \right] $ \\[6pt]
BIC & $\displaystyle \max\left( 0, \frac{LL-LL_0}{N} -\frac{K\log N}{2N} \right) $ & $\displaystyle \left[ 0, \log J-\frac{K\log N}{2N} \right] $ \\[6pt]
McFadden's pseudo-$R^2$ & $\displaystyle \max\left( 0, 1-\frac{LL}{LL_0} \right) $ & $\displaystyle [0,1] $ \\[6pt]
Adjusted McFadden's pseudo-$R^2$ & $\displaystyle \max\left( 0, 1-\frac{LL-K}{LL_0} \right) $ & $\displaystyle \left[ 0, 1+\frac{K}{N\log J} \right] $ \\
\bottomrule
\end{tabularx}
\end{table}

For log-likelihood, Delphos measures the average improvement in model fit per observation relative to the null specification, and it is bounded by $0 \leq LL \leq -LL_0/N$. For information criteria, the direction of improvement is reversed because lower values indicate preferred specifications. We also divide each criterion by divided by $2N$ to express both metrics on the same per-observation log-likelihood scale. Accordingly, the AIC-based reward is bounded by $\left[0,\log J-K/N\right]$, while the BIC-based reward is bounded by $\left[0,\log J-K\log N/(2N)\right]$. Finally, McFadden's pseudo-$R^2$ measures require no additional scaling because they are already defined relative to a reference log-likelihood.\\

 This flexible design allows the reward function to reflect different profiles of choice modellers. A modeller focused on model fit may emphasise log-likelihood or pseudo-$R^2$, while one concerned with complexity may prioritise information criteria or a balance between model fit and number of parameters. When behavioural plausibility is also required, penalties for behavioural violations guide the agent not only towards well-performing models, but also theoretically consistent ones.  Finally, future work could improve this framework by incorporating other domain-specific constraints. For instance, hypothesis testing against the null could be used to penalise model specifications in which most parameters are statistically insignificant. \\

\subsubsection{The agent}\label{sec: agent}
\begin{figure}[ht]
    \centering
     \includegraphics[width=0.9\linewidth]{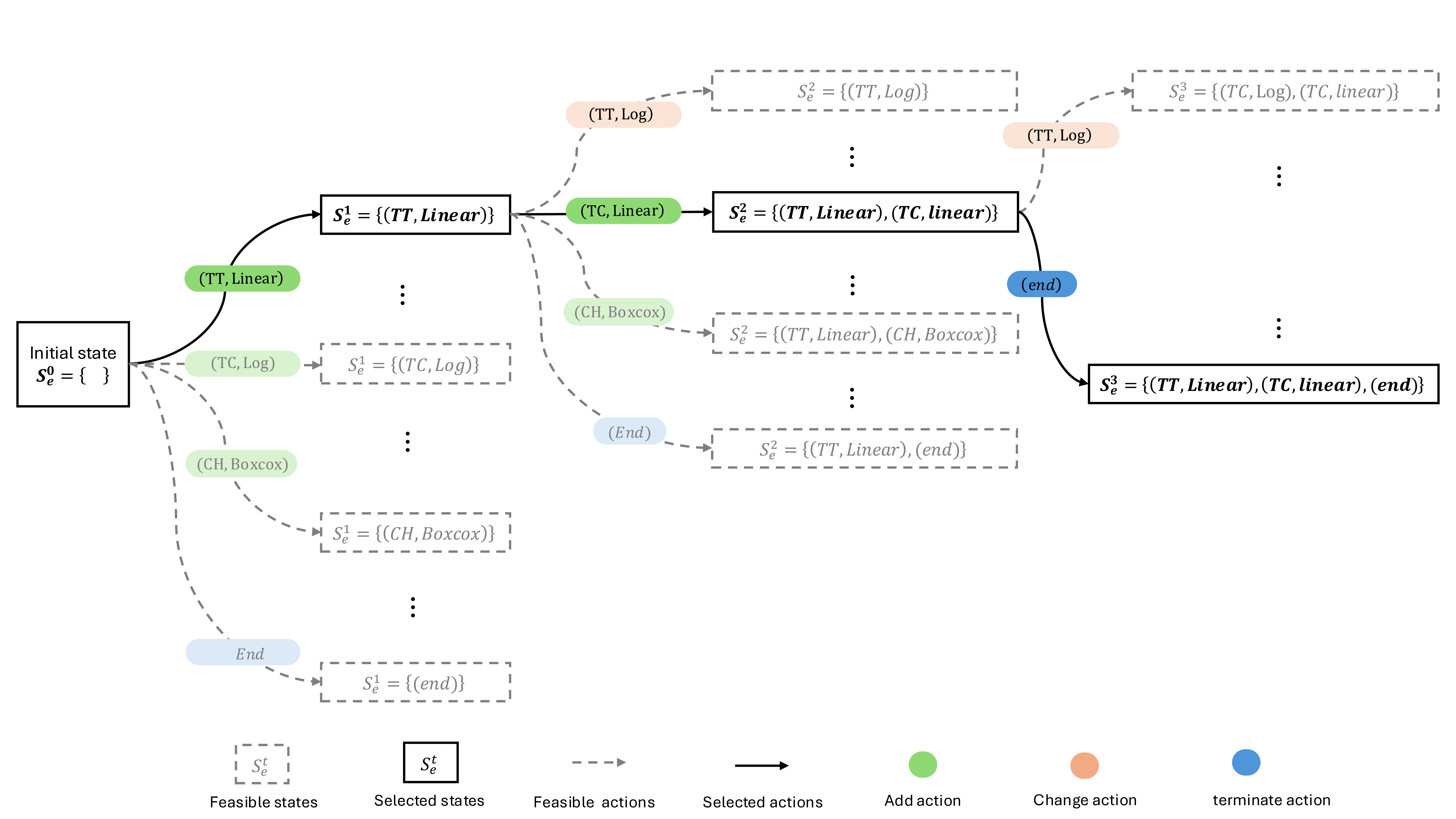}
    \caption{Agent’s sequential decision-making process.
    Nodes represent internal states (current specifications). Dashed arrows indicate feasible actions, while solid arrows show the actions selected by the agent until termination.}
    \label{fig: agent_sequential_DM}
\end{figure}

Delphos follows a DQN architecture, enabling it to learn a policy that selects actions based on their Q-values through interaction with the environment \citep{mnih2015human}. As displayed in Figure~\ref{fig: DQN_DCM_framework}, Delphos encodes the current model specification into a fixed-length state representation, uses a DQN to approximate the state--action value function, selects actions accordingly, interacts with the environment to estimate the specified models, and stores its experiences in the replay buffer for training. To illustrate this process, consider a transport mode choice scenario involving different alternatives and attributes: travel time (TT), travel cost (TC), headway (HW), and number of changes (CH). In this example, Delphos’ objective is to specify a choice model with fully generic taste parameters, as shown in Figure \ref{fig: agent_sequential_DM}.\\

Delphos begins each episode with an empty fixed-length  state $(\,S_e^0  = \{\, (k_q,0,0,0)\,\}_{q=1}^{K+1}\,)$, which represents an unspecified model. As the episode progresses, the agent updates the internal state $(S_e^l)$ by selecting attributes, applying transformations, and incorporating interactions.  At each intermediate step, the current specification is encoded as a fixed-length multi-hot vector $\mathbf{s}_e^l$, with one binary entry for each feasible combination of attribute, functional form, taste parameter structure, and covariate interaction. Because these entries are determined by the modelling space rather than the sequence of actions, identical specifications yield the same order-invariant representation. The encoded state is then passed through the input layer of the DQN, where the policy network processes it using fully connected hidden layers. The output layer provides Q-values for every action, while the feasibility mask is subsequently applied to exclude invalid actions. Finally, the agent selects an action according to its $\epsilon$-greedy policy.\\

Once Delphos chooses the terminal action, the environment estimates the candidate model and returns the corresponding modelling outcomes, from which the episodic reward is calculated. If the candidate has already been estimated, its stored outcomes are retrieved instead of estimating it again. These resulting trajectories (the internal states, selected actions, intermediate discounted returns, and subsequent internal states) are stored as transitions  $(S_e^l, a_e^l, G_e^l, S_e^{l+1})$ in a replay buffer, which enables the agent to learn based on past modelling outcomes as it refines its policy. To do so, Delphos samples mini-batches from this buffer to update the policy network parameters, thereby refining its decision-making over time, as described in section~\ref{sec: RL}.\\

While Delphos is designed to learn a policy for sequential model specification through iterative interactions, it can also be adapted to automate the search for high-performing specifications with a limited computational budget. When the aim is to train a stable policy, the agent is trained for a predefined horizon $E_{\max}$, and its policy stability is assessed from its performance over the final training period. In contrast, when the aim is to identify a high-performing specification within a limited number of episodes, an optional early-stopping rule terminates training when the best-so-far reward has ceased to improve. Formally, after a minimum of $E_{\min}$ episodes, training terminates if $R_e^\star-R_{e-E_{\text{search}}}^\star\leq\delta_R$, where $R_e^\star=\max_{1\leq j\leq e}R_j$ denotes the highest reward observed by Delphos up to episode $e$, $E_{\text{search}}$ is the patience window, and $\delta_R$ is the minimum improvement considered meaningful. The predefined horizon training remains an upper bound if the early-stopping condition is not met. This criterion reduces computation when further exploration does not produce better specifications, although it does not provide evidence of policy convergence.

\subsubsection{The environment}\label{sec: environment}
The environment handles model estimation and returns the associated modelling outcomes. It is initialised with a choice dataset and remains fixed throughout all agent interactions. During each episode, the environment receives encoded state representations from the agent, decodes them into utility specifications, performs the estimation process, and returns the modelling outcomes required to calculate the reward signal. Built using Apollo functions \citep{hess2019apollo}, the environment meets the OpenAI Gym requirements \citep{brockman2016openai} that ensure compatibility with a range of RL algorithms. Importantly, the reward function is deliberately isolated from the environment. This design provides modellers with flexibility to define custom reward signals, including the integration of behavioural constraints (e.g., penalising unexpected coefficient signs) and the use of multiple modelling outcomes, as discussed previously. We have made the code public in a GitHub repository: \url{https://github.com/TUD-CityAI-Lab/Delphos-single-task}

\section{Experiments}\label{sec: monte carlo exp}

To evaluate the capabilities of Delphos, we design a sequence of experiments that progress from simulated scenarios to empirical cases. The simulated experiments assess the agent’s ability to recover known data-generating processes, navigate modelling spaces, and learn a stable policy that yields high rewards. We then benchmark Delphos against a Variable Neighbourhood Search algorithm (VNS) \cite{ortelli2021assisted} and Simulated Annealing (SA) \cite{paz2019specification, ahern2026searchlibrium} using the Swissmetro dataset \citep{bierlaire2001acceptance} in a realistic empirical setting where model performance and computational efficiency are relevant.  Finally, we apply Delphos to the DECISIONS dataset \citep{calastri2020we} to evaluate whether a modeller-driven reward that combines model fit, parsimony, and behavioural plausibility can guide the agent to propose competitive specifications that satisfy these constraints set by the analyst. Across all experiments, each dataset is split into in-sample (80\%) and out-of-sample (20\%) subsets. The agent is trained solely on the in-sample set, where it iteratively explores the modelling space and refines its decision-making process (policy). Once training is completed, the resulting candidate specifications are evaluated out-of-sample to assess generalisability to unseen data.\\

We use multiple experimental configurations that vary in reward formulation and early stopping criteria. These configurations serve two complementary purposes. First, they allow us to reflect the diversity of modeller goals observed in empirical choice modelling workflows \citep{nova2025understanding}, including preferences for model fit, parsimony, and behavioural expectations. Second, they enable us to investigate the agent’s ability to identify competitive specifications within a restricted number of episodes, rather than requiring full policy convergence. Overall, these settings allow us to evaluate how effectively the agent balances exploration and exploitation, to assess the robustness of its learned policy across simulated datasets, and to propose competitive models under different modelling goals and computational constraints in empirical cases.\\

\subsection{Experiment 1: Simulated datasets}
We first evaluate Delphos in controlled conditions using three simulated choice datasets based on Multinomial Logit models. Each dataset contains 4,000 individuals choosing among 3  unlabelled alternatives, defined by $K$ predefined attributes ($x_1, x_2, ..., x_k$). We also simulate $C$ socio-demographic variables (cov$_1$,..., cov$_C$) to introduce observed systematic heterogeneity.  Table \ref{tab:true-specifications} shows the true utility specifications\footnote{True parameter values are provided in Appendix~\ref{appendix: True_parameters}.} and the modelling space available to the agent for each case. In all cases, the agent is trained using the adjusted McFadden’s pseudo-$R^2$ as the reward signal. Additionally, learning parameters are varied to evaluate the sensitivity of the agent’s performance to different hyperparameter settings (Appendix~\ref{appendix: hyperparameter}). This setup enables us to investigate whether the agent is effectively learning,  whether its policy concentrates on high-reward regions of the modelling space, and whether it is capable of recovering the true data-generating process under different levels of complexity.

    \begin{table}[ht]
    \centering
    \caption{Utility specifications and modelling spaces for simulated choice datasets ($1A$--$1C$).}
    \label{tab:true-specifications}
    \resizebox{0.95\textwidth}{!}{
    \begin{tabular}{c|lc|lc|lcc}
    \toprule
     \textbf{Experiment}
     & \multicolumn{2}{c|}{\textbf{$1A$}}
     & \multicolumn{2}{c|}{\textbf{$1B$}}
     & \multicolumn{3}{c}{\textbf{$1C$}} \\\hline
    \textbf{Variable}
     & \multicolumn{1}{c}{Transf.} 
     & \multicolumn{1}{c|}{Taste} 
     & \multicolumn{1}{c}{Transf.} 
     & \multicolumn{1}{c|}{Taste} 
     & \multicolumn{1}{c}{Transf.} 
     & \multicolumn{1}{c}{Taste} 
     & \multicolumn{1}{c}{Covariate} \\
    \midrule 
    ASC$_1$ & Linear &              & Linear &            &Linear  &  & cov$_1$ \\
    ASC$_2$ & Linear &              & Linear &            & Linear &  & cov$_1$ \\
    $x_1$   & Linear & Generic      & Log & Specific                & Linear & Specific & \\
    $x_2$   & Log    & Generic      & Box-Cox & Generic             & Log & Specific & cov$_2$ \\
    $x_3$   & Linear & Generic      & Linear & Specific         & Linear & Generic & \\
    $x_4$   & Box-Cox & Generic     & Linear & Generic             & Box-Cox & Generic & \\
    $x_5$   &Linear & Generic       &  &  &             Linear & Generic & \\
    $x_6$   &None &        &  &  &              &  & \\
    \midrule
    Space
     & \begin{tabular}{@{}l@{}}None\\Linear \\ Log \\ Box-Cox\end{tabular}
     & \begin{tabular}{@{}l@{}}Generic\end{tabular}
     & \begin{tabular}{@{}l@{}}None\\Linear \\ Log \\ Box-Cox\end{tabular}
     & \begin{tabular}{@{}l@{}}Generic \\ Specific\end{tabular}
     & \begin{tabular}{@{}l@{}}None\\Linear \\ Log \\ Box-Cox\end{tabular}
     & \begin{tabular}{@{}l@{}}Generic \\ Specific\end{tabular}
     & \begin{tabular}{@{}l@{}}None\\cov$_1$ \\ cov$_2$ \end{tabular} \\
    \bottomrule
    \end{tabular}}
    \end{table}

\subsection{Experiment 2: Empirical datasets}
\subsubsection{Swissmetro}
We next analyse Delphos in an empirical case using the Swissmetro dataset, a stated preference dataset designed to evaluate the introduction of a new high-speed public transport system (for more details, see \cite{antonini2007swissmetro}).  The dataset contains 1,192 respondents, each facing nine hypothetical mode choice tasks among train, swissmetro, and car. Alternatives are defined by travel time (TT), travel cost (TC), headway (HE), and seat configuration (SE). It also includes socio-demographic variables such as age, income, travel class, possession of an annual season ticket (GA), number of suitcases (LUGGAGE), gender (GENDER), and the person or employer responsible for paying the ticket (WHO). \\

For this experiment, we  benchmark the specifications proposed by Delphos against those obtained by VNS and SA in terms of AIC. To ensure a fair and robust comparison while accounting for stochasticity, all methods are evaluated under identical experimental conditions and run across 10 independent seeds\footnote{We run Delphos using the hyperparameters identified in Experiment 1C (Appendix~\ref{appendix: hyperparameter})}. In particular, we use the same in-sample dataset, the null model as the starting specification, and the same budget of model estimations to terminate the search process ($B=1,500$). Importantly, we restrict all methods to the same modelling space, which accommodates non-linear transformations, generic and alternative-specific taste parameters, and interactions with socio-demographic variables. This design allows us to benchmark Delphos directly against metaheuristics while assessing whether the agent can propose competitive AIC models within a constrained number of episodes.\\

\subsubsection{DECISIONS}
Finally, we evaluate Delphos empirically using the DECISIONS dataset (for more details, see \cite{calastri2020we}), focusing on its ability to accommodate modeller-defined objectives and explicitly incorporate behavioural plausibility into the reward function. This dataset integrates an online survey and a two-week mobile travel diary, capturing information on travel behaviour, residential location, energy use, and social network interactions. It includes alternative-specific attributes—such as in-vehicle travel time (IVT), out-of-vehicle travel time (OVT), and total travel cost (TC)—and covariates such as trip purpose, city, income, gender, age, education, and household size.\\  

In this experiment, the agent can decide on attribute transformation, specify generic or alternative-specific tastes,  and interactions with covariates, as in the previous experiment. The reward function is designed to represent an analyst concerned with model performance and model parsimony while enforcing behavioural expectations. It thus not only combines the log-likelihood component (weight 0.7) with a penalty on the number of parameters (weight 0.3), but also incorporates behavioural expectations by requiring negative estimated parameters for in-vehicle travel time and total travel cost. To automate the search within a constrained budget, we employ an early stopping criterion, the search terminates if the agent fails to identify candidates that improve the reward beyond a 0.001 tolerance over a patience window of 2,000 model attempts. This final experiment evaluates whether Delphos can propose specifications that are both well-performing and theoretically consistent when guided by a modeller-like reward formulation under constrained learning conditions.    

\section{Results}\label{sec: results}
We present the results of the experimental evaluation of Delphos in two parts. We first analyse the agent’s learning behaviour in simulated environments, where training runs until the policy stabilises, allowing us to evaluate convergence properties, exploration efficiency, and the ability to approach the true data-generating process (Subsection~\ref{sec:simulated_experiments}). We then examine the empirical applications (i.e. experiments 2 and 3), where the focus shifts towards identifying well-performing and behaviourally plausible specifications within constrained learning horizons (Subsection \ref{sec:empirical_experiments}), including a direct benchmark against a metaheuristic on Swissmetro and an analyst-driven specification task on DECISIONS.

\subsection{Results simulated experiments}\label{sec:simulated_experiments}
The simulated experiments evaluate whether Delphos learns a stable specification strategy when the true data-generating process is known. We analyse (i) learning curves of episodic rewards that capture how the agent’s episodic reward evolves over time, (ii) the evolution of Q-values across training to understand how the agent prioritises actions throughout training, and (iii) measure exploration coverage metrics that reveal how efficiently the agent explores the modelling space and concentrates on high-reward regions. Results are averaged over ten independent runs with different random seeds.\\

\begin{figure}[ht]
    \centering
    \includegraphics[width=0.75\linewidth]{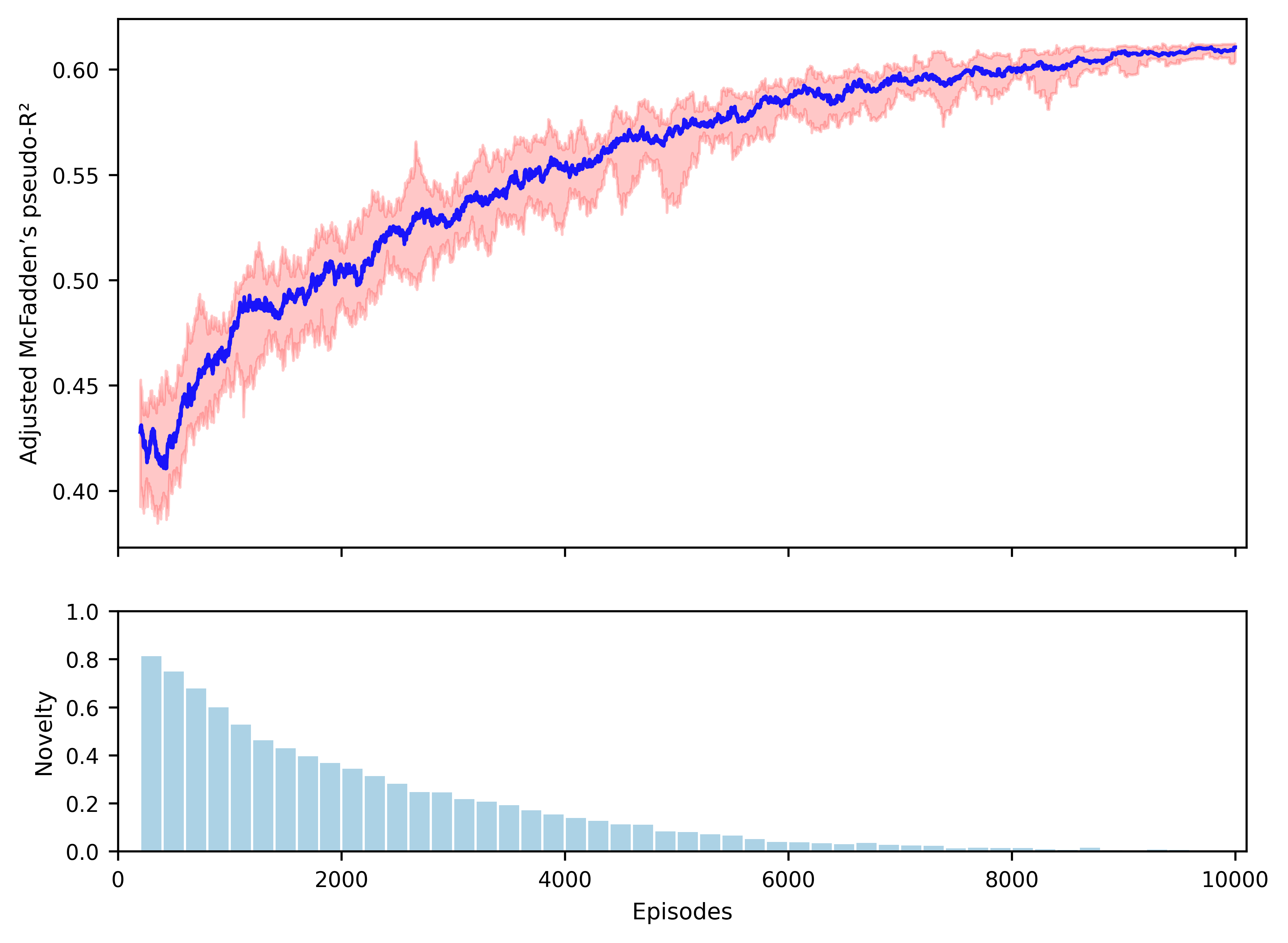}
    \caption{Learning process on simulated dataset $1A$. The upper panel displays the learning curve, showing the rolling mean and min–max range of the reward signal. The lower panel reports rolling novelty, measured as the share of previously unseen specifications.}
    \label{fig:exp1a}
\end{figure}

\noindent Figure~\ref{fig:exp1a} shows the RL agent’s learning process and novelty over episodes, considering the simulated choice dataset $1A$. The upper figure shows the rolling mean and the min-max range of the learning curve across ten runs with different random seeds, using $\bar{\rho}^2$ as the reward signal over $10,000$ training episodes. Initially, the agent’s learning trajectory starts at low reward levels, which reflect an exploratory phase where random actions are taken to specify candidate models. As training progresses, the learning curve gradually rises until it stabilises, which indicates a shift from exploration to exploitation, where the agent prioritises actions that have previously yielded higher reward signals. The upward trend suggests that the agent improves its policy and eventually stabilises on modelling decisions that yield well-performing specifications. The spread across seeds also narrows over time, which indicates consistent performance towards the end of training.

\noindent The lower figure computes novelty as the share of previously non-proposed specifications within the same rolling window and averaged across runs. This value is highest early in training and then declines towards zero, indicating that the agent progressively shifts from proposing new specifications (exploration) to revisiting a smaller set of candidates (exploitation). Because this drop coincides with improving and stabilising rewards, it suggests that the agent's learning is driven by policy refinement rather than continued exploration. Moreover, Delphos estimates 1{,}890 unique models across 10{,}000 episodes, implying that it revisits previously seen specifications in many episodes without re-estimating them, while the agent refines its policy.\\

\begin{figure}[ht]
    \centering
    \includegraphics[width=0.75\linewidth]{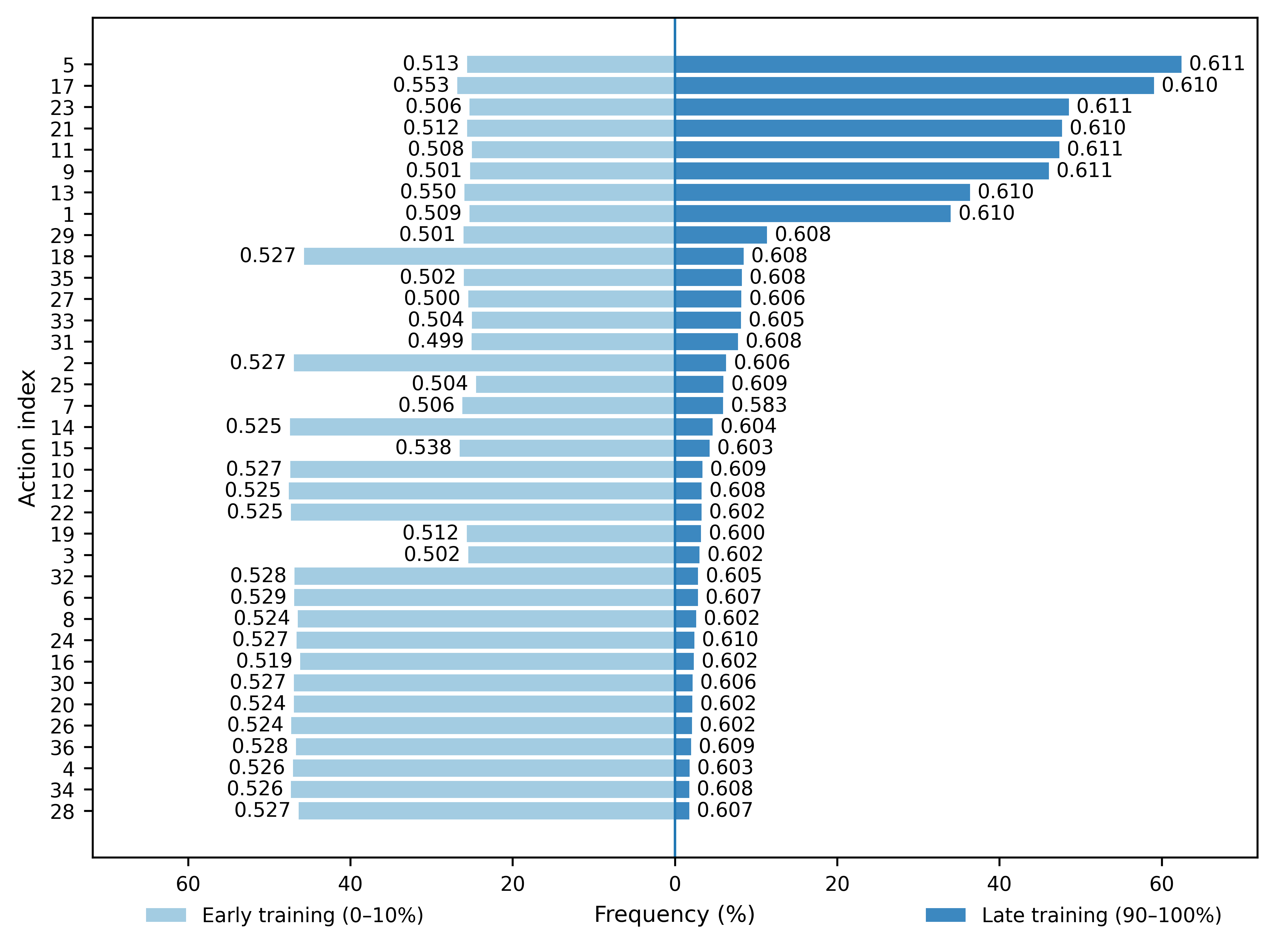}
    \caption{Action selection distribution in simulated dataset $1A$. Bars show the average frequency with which each action is selected in the early and late training across runs. Numbers report the average $\bar{\rho}^2$ for specifications in which the action was used. The full list of actions and their corresponding definitions is provided in the Appendix~\ref{appendix:action-space}.}
    \label{fig:exp1c}
\end{figure}

\noindent To analyse the agent's policy, Figure~\ref{fig:exp1c} shows how the agent’s modelling actions change from the start to the end of the training process. For each modelling action, we display the average frequency with which it is taken in the early and late training stages across runs, ranked by its late-stage frequency. We also report the mean adjusted McFadden’s pseudo-$R^2$ obtained in the specifications where that action was used. In the early episodes, the agent explores the modelling space by selecting a broad and relatively uniform set of actions, resulting in dispersed Q-values. As training progresses, the policy is updated, and Q-values increasingly concentrate around a smaller subset of actions. At the end of the training process, the agent relies on a smaller subset of actions that yield high-reward specifications, which indicates that it has become more selective in how it adjusts the model specification. This is further supported by the fact that some actions that remain low-frequency in late training still achieve high mean rewards when they are used. This suggests they are applied only in certain states (internal steps) where they are effective, rather than uniformly in the early exploratory phase.\\ 

\noindent Compared to Experiment $1A$, the learning curves in Experiments $1B$ and $1C$ show higher variability in the min-max range across the ten runs (additional figures are provided in Appendix~\ref{appendix: simulated experiment 1B and 1C}). This increased noise reflects the added complexity of incorporating covariates into the model specification process. Covariates expand the modelling space substantially, making it more challenging for the agent to consistently identify high-performing regions, and they introduce additional opportunities for misspecification, leading to more diverse reward signals. Consequently, it takes longer for the agent to recognise which sequences of actions are valuable to take in future episodes for refining its policy. The inclusion of covariates significantly enlarges the action space, making the Q-value distribution more diffuse. Nonetheless, the learning curve eventually reaches a plateau, which indicates that the agent converges towards a stable policy that proposes high-performing model specifications despite the increased complexity of the task.\\

Table \ref{tab:occupancy-metrics} summarises the modelling space parameters, exploration coverage, and performance for each simulated case. We report the size of the modelling space (i.e., the total number of feasible specifications), the number of unique specifications explored by the agent, the percentage of successfully converged estimations, and the episode at which the true specification is first recovered. Finally, we report the average reward over the last 100 episodes ($\bar{\rho}_{100}$), alongside its gap relative to the true specification's reward ($\Delta \text{Reward} = \bar{\rho}_{100} - \rho_{true})$.\\


\begin{table}[ht]
\centering
\caption{Modelling size, exploration coverage, and performance on simulated experiments}
\label{tab:occupancy-metrics}
\resizebox{\linewidth}{!}{%
\begin{tabular}{lcccccc}
\toprule
\textbf{Exp} & \textbf{Model space} & \textbf{Models estimated by agent} & \textbf{Converged models} & \textbf{First recovery} & \textbf{Reward (last 100)} & \textbf{$\Delta$Reward} \\
\midrule
1A & 8.19 $\times 10^3$ & 1.89 $\times 10^{3}$ & 97.42\% & 1.46 $\times 10^{3}$ & 0.611 & -0.002 \\
1B & 8.19 $\times 10^3$ & 2.34 $\times 10^{3}$ & 98.45\% & 1.63 $\times 10^{3}$ & 0.753 & -0.009 \\
1C & 1.59 $\times 10^6$ & 3.92 $\times 10^{3}$ & 78.48\% & 3.73 $\times 10^{3}$ & 0.654 & -0.023 \\
\bottomrule
\end{tabular}}
\end{table}

\noindent As expected, the modelling space in Experiment $1C$ is orders of magnitude larger than in Experiments $1A$ and $1B$, due to the inclusion of covariates and their associated interactions. Across all experiments, the agent does not attempt to exhaustively explore the entire modelling space but instead concentrates its search on a comparatively small subset of candidates. In Experiment $1A$, the agent explored $1,899$ unique models and recovered the true specification by episode $1,461$ on average, achieving a $97.42$\% convergence rate. In late training, the learned policy concentrates on high-performing candidates whose rewards remain close to the true model value ($\Delta\bar{\rho}^2=-0.001$ over the last 100 episodes), which indicates a stable near-optimal performance. A similar agent behaviour is observed in Experiment $1B$, where it explored $2,338$ unique models, identified the true specification by episode $1,627$, achieved a $98.45$\% convergence rate, and the final policy again yields near-optimal specifications ($\Delta\bar{\rho}^2=-0.009$). \\

\noindent In contrast, Experiment $1C$ involved navigating an exponentially larger modelling space. The agent explored $3,922$ unique specifications (less than 1\% of the feasible set), and the convergence rate drops to $78.48$\%. Although the true specification is not recovered across runs, the final learnt policy still concentrates on near-optimal candidates; the mean rewards over the last 100 episodes remain close to the true model value ($\Delta\bar{\rho}^2=-0.023$). The agent also reaches its best observed reward much before the training horizon (on average at episode $3,731$), and this maximum reward ($0.675$) remained close to the reward value obtained by the true model (see Appendix~\ref{appendix: simulated experiment 1B and 1C}). This suggests that Delphos can learn a policy that identifies well-performing candidates even in high-dimensional and combinatorial spaces.  Importantly, for real-world applications where the true data-generating process is unknown, the goal is not exact recovery but the agent’s ability to consistently converge towards near-optimal modelling regions within such environments.

\subsection{Empirical experiments}\label{sec:empirical_experiments}
The empirical experiments evaluate Delphos under realistic and high-dimensional modelling spaces. The agent’s objective is not to learn a fully stable policy but to identify competitive specifications within a constrained budget of estimation models. We first benchmark Delphos against two metaheuristics (VNS and SA) on the Swissmetro dataset. We then run Delphos on the Decision datasets to analyse whether the agent can propose models that satisfy an modeller-defined reward formulation combining parsimony and behavioural constraints.

\subsubsection{Benchmarking on Swissmetro}
We first evaluate Delphos against SA and VNS in their capacity to identify competitive model specifications on the Swissmetro dataset, using AIC as the reward function for Delphos and the cost function for the metaheuristics. To ensure robustness, we run 10 independent seeds across each method\footnote{Experiments were run on a 4.1 GHz 12-core Apple M3 Pro processor with 18 GB of unified memory}. All of them start from a null model, operate within the same modelling space of approximately 88.5 million possible specifications, but constrain their search to a budget of 1,500 model estimation attempts per run.

\begin{table}[ht] 
\centering 
\caption{Search space performance across 10 independent seeds.} 
\label{tab:search_metrics} 
\resizebox{\textwidth}{!}{
\begin{tabular}{l ccc ccc cc} 
\toprule
\multirow{2}{*}{} & \multicolumn{3}{c}{Search Metrics} & \multicolumn{3}{c}{In-Sample AIC} & \multicolumn{2}{c}{Out-of-Sample LL} \\
\cmidrule(lr){2-4} \cmidrule(lr){5-7} \cmidrule(lr){8-9}
        & \#Param. & Converged & Time (min) & Mean ($\pm$ std) & Best & Worst & Best & Worst \\ 
\midrule 
Delphos & 22 ($\pm$8) & 645 ($\pm$65)   & 24 ($\pm$3)   & 7,957 ($\pm$394) & 7,454 & 9,128 & -879 & -1,073 \\ 
SA      & 13 ($\pm$7) & 1,344 ($\pm$21) & 9 ($\pm$2)    & 8,431 ($\pm$472) & 7,463 & 9,410 & -913 & -1,163 \\ 
VNS     & 8 ($\pm$4)  & 1,443 ($\pm$14) & 104 ($\pm$23) & 8,843 ($\pm$385) & 7,744 & 9,401 & -948 & -1,163 \\ 
\bottomrule 
\end{tabular}
}
\end{table}

First, table~\ref{tab:search_metrics} summarises the search metrics and both in-sample and out-of-sample performance across seeds for each method. In terms of search performance, Delphos explored an average of 645 successfully converged model specifications per seed. The low convergence rate indicates that the agent's policy proposes specifications that are more complex by including non-linear transformations and socio-demographic variables in early phase. Meanwhile, SA and VNS successfully converged an average of 1,344 and 1,443 models, respectively, as they explored simpler specifications. Regarding in-sample performance, Delphos achieves the lowest mean AIC (7,957 $\pm$ 394) alongside the single best specification across all seeds (AIC = 7,454), outperforming SA (mean AIC = 8,431; best = 7,463) and VNS (mean AIC = 8,843; best = 7,744). Importantly, the standard deviation of Delphos across seeds ($\pm$ 394) confirms that its learned policy consistently guides the search toward high-performing regions regardless of initialisation. Moreover, the average of number of parameters for Delphos highlights its tendency to navigate complex spaces compared to the metaheuristics. Finally, when evaluating computational speed, SA is the fastest algorithm (9 minutes) due to its JAX-based backend (SearchLibrium 0.0.251), followed by Delphos (24 minutes) using Apollo (0.3.7), and VNS (104 minutes) using Biogeme (3.3.1). To evaluate generalisation, we took the best and worst specifications using the in-sample AIC and compute their out-of-sample log-likelihood on the held-out set. Delphos's best in-sample specification achieves the highest out-of-sample log-likelihood (OOS LL = -879; LL/N = -0.65), which outperforms the best specifications of SA (OOS LL = -913; LL/N = -0.67) and VNS (OOS LL = -948; LL/N = -0,70). 

\begin{figure}[ht]
    \centering
     \includegraphics[width=0.8\linewidth]{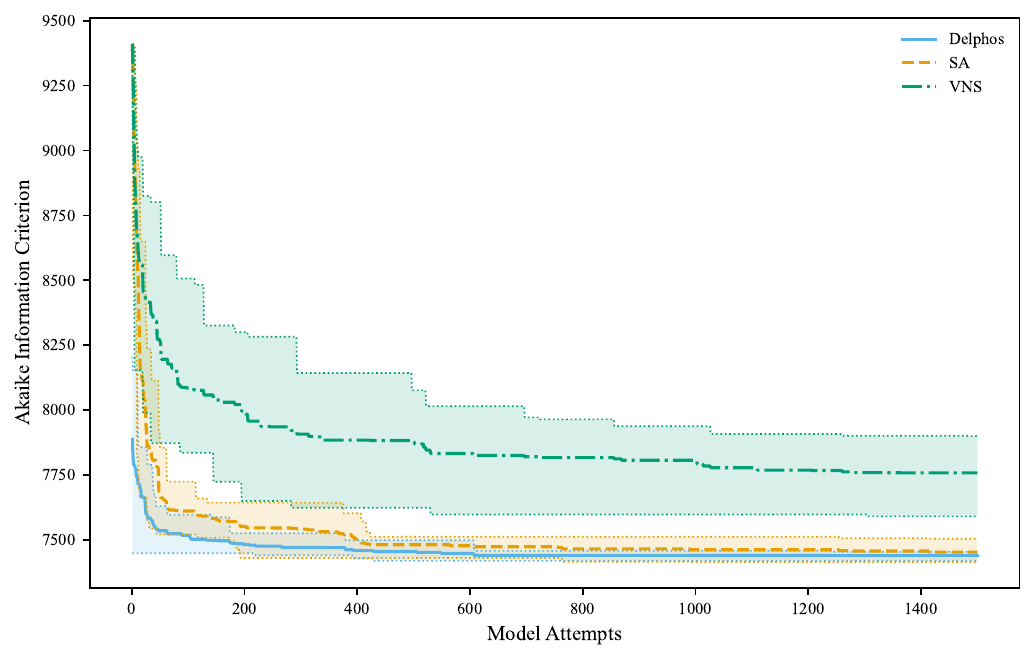}
    \caption{Evolution of the best-so-far AIC  across methods on the Swissmetro dataset.}
    \label{fig: exp_swissmetro_best}
\end{figure}

 Second, we track the best-performing models across estimation attempts using the raw modelling  outcome (AIC), as shown in Figure~\ref{fig: exp_swissmetro_best}. This plot reports the fastest, slowest, and mean best-so-far curves across seeds, reflecting how each method incrementally proposes candidates that outperform previous ones. The rapid improvement during the initial model attempts suggests that Delphos is able to identify promising candidates, reaching a mean performance near an AIC of 7,500 within the first 130 model attempts and achieving its minimum (7,439) at model attempt 1,239. While SA requires nearly 400 attempts to approach a comparable performance, eventually reaching a minimum mean AIC of 7,452 at attempt 1414. VNS shows the slowest progression, which fails to converge to the same performance level within the allotted budget and stalls at a mean minimum of 7,758. This behaviour highlights the fundamental difference between the search algorithms. Because Delphos formulates the specification process as a sequence of modelling decisions over partial utilities, it can immediately construct complex specifications by accommodating non-linearities and observed heterogeneity in a single model attempt. Importantly, it achieves this process without requiring local movements from previously evaluated candidates. In contrast, the SA and VNS metaheuristics rely on local search, which must incrementally build upon the null model and past candidates through local operations. Delphos can, therefore, associate rewards with actions that are prone to yield high-performing specifications, without relying on candidate trajectory. \\

\begin{figure}[ht]
    \centering
     \includegraphics[width=0.8\linewidth]{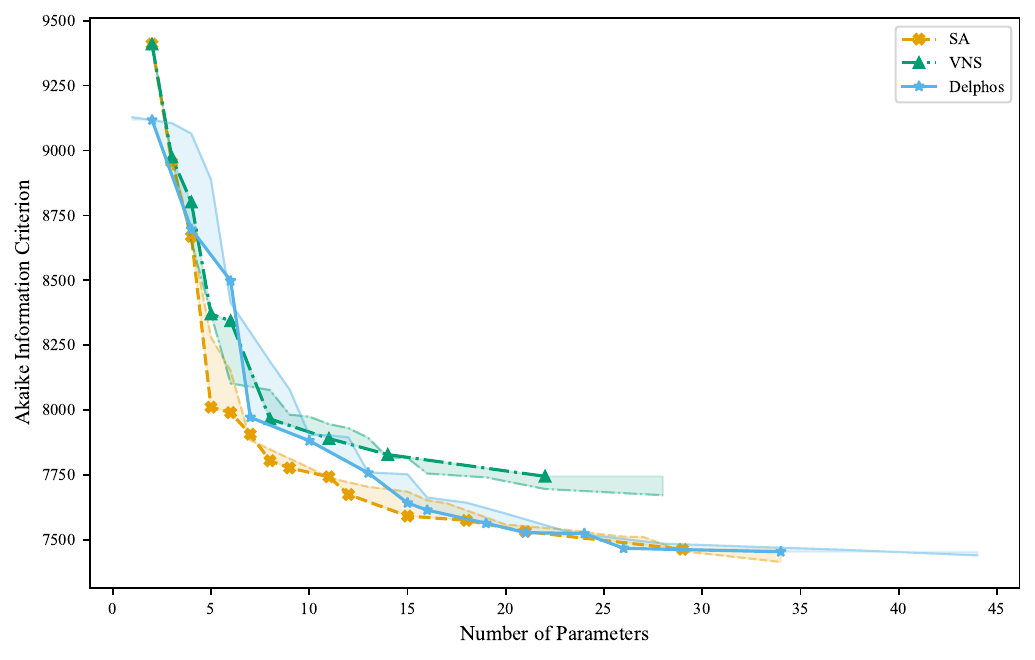}
    \caption{Pareto front of  Delphos, SA, VNS on the Swissmetro dataset.Bold lines indicate the Pareto front of the best-performing seed for each method, while shaded regions span the area between the best and worst seeds across the 10 independent runs.}    
     \label{fig: exp_swissmetro_pareto}
\end{figure}

To further evaluate the trade-off between competing modelling objectives, we built the Pareto front of  successfully estimated models that represent an optimal balance between model performance (AIC) and parsimony (Number of parameters) across seeds. Figure~\ref{fig: exp_swissmetro_pareto} shows the Pareto fronts for the best- and worst-performing seed per method, with shaded regions representing their variance. Given the estimation budget, SA tends to dominate the low-complexity range (models with fewer than 15 parameters), where its local search efficiently explores the space near the null model. VNS yields a compact Pareto front and stalls before 20 parameters, suggesting that it requires a larger number of model evaluations to navigate higher model complexity. In the mid-complexity range (15 to 25 parameters), SA and Delphos overlap almost completely, reaching comparable performance. However, as complexity increases beyond 30 parameters, Delphos extends the frontier past 40 parameters, outperforms the metaheuristics, and identifies the lowest AIC specification. Although Delphos shows a wider variance in the low-complexity, its variance narrows significantly towards higher complexity. This suggests that the independent seeds converge to similar AIC values. The modelling results for the lowest in-sample AIC specification, including estimate values and standard errors, can be found in Table~\ref{tab:estimates_swissmetro_delphos}. \\

Overall, this benchmark suggests that all three approaches identify competitive specifications under the same experimental setup, with VNS and SA tending to propose models with fewer parameters, while Delphos identifies lower-AIC specifications at higher levels of model complexity. This difference is consistent with the fact that Delphos updates its action policy using the reward signal obtained after each estimation, where only successfully converged models receive positive feedback. As a result, the agent can continuously move away from those actions that frequently lead to non-convergence while continuing to explore higher-complexity regions. By contrast, VNS and SA treat non-convergent candidates as invalid and exclude them from the Pareto set in post-processing, rather than using these outcomes to learn a decision strategy that guides subsequent modelling decisions.

\subsubsection{Modeller-driven reward design with behavioural constraints on DECISIONS}
We next apply the agent to the DECISIONS dataset to evaluate its ability to suggest model specifications under knowledge-driven learning objectives. Specifically, we consider: (i) a weighted sum of normalised log-likelihood ($\widetilde{LL}$) and the normalised number of parameters ($\widetilde{K}$) as the reward signal; and (ii) the incorporation of behavioural expectations, such as sign constraints on specific parameters. Formally, the episodic reward is $R_e = \mathds{1}_{\text{converged}} \cdot \mathds{1}_{\text{time}}\cdot \mathds{1}_{\text{cost}} \cdot \big (0.8 \cdot  \widetilde{LL} + 0.2 \cdot \widetilde{K}\big )$. In this setting,  Delphos explored 2,660 unique candidates over the 6.72 million feasible model specifications in the modelling space, requiring approximately 7.4 hours (an average of 13.52 seconds per model estimation). Across these converged candidates, model complexity ranged from $2$ to $85$ parameters, and the log-likelihood ranged from $-11,782.73$ to $-3,854.37$.\\

\noindent Delphos identifies a high-performing specification balancing goodness-of-fit and parsimony as defined by the reward function. The best LL-NumParam candidate satisfies the behavioural constraints encoded in the reward design, including negative coefficients for in-vehicle travel time and travel cost. As training progresses, the agent no longer identifies candidates that improve the modeller-defined reward, and early stopping is triggered once additional episodes fail to yield further improvements. Table~\ref{tab:estimates_decisions_delphos} in the Appendix reports the modelling outcomes for the best LL–NumParm specification identified by Delphos, including parameter estimates, robust standard errors, and coefficient signs.

 \begin{figure}[h]
    \centering
    \vspace{1em}
     \includegraphics[width=0.8\linewidth]{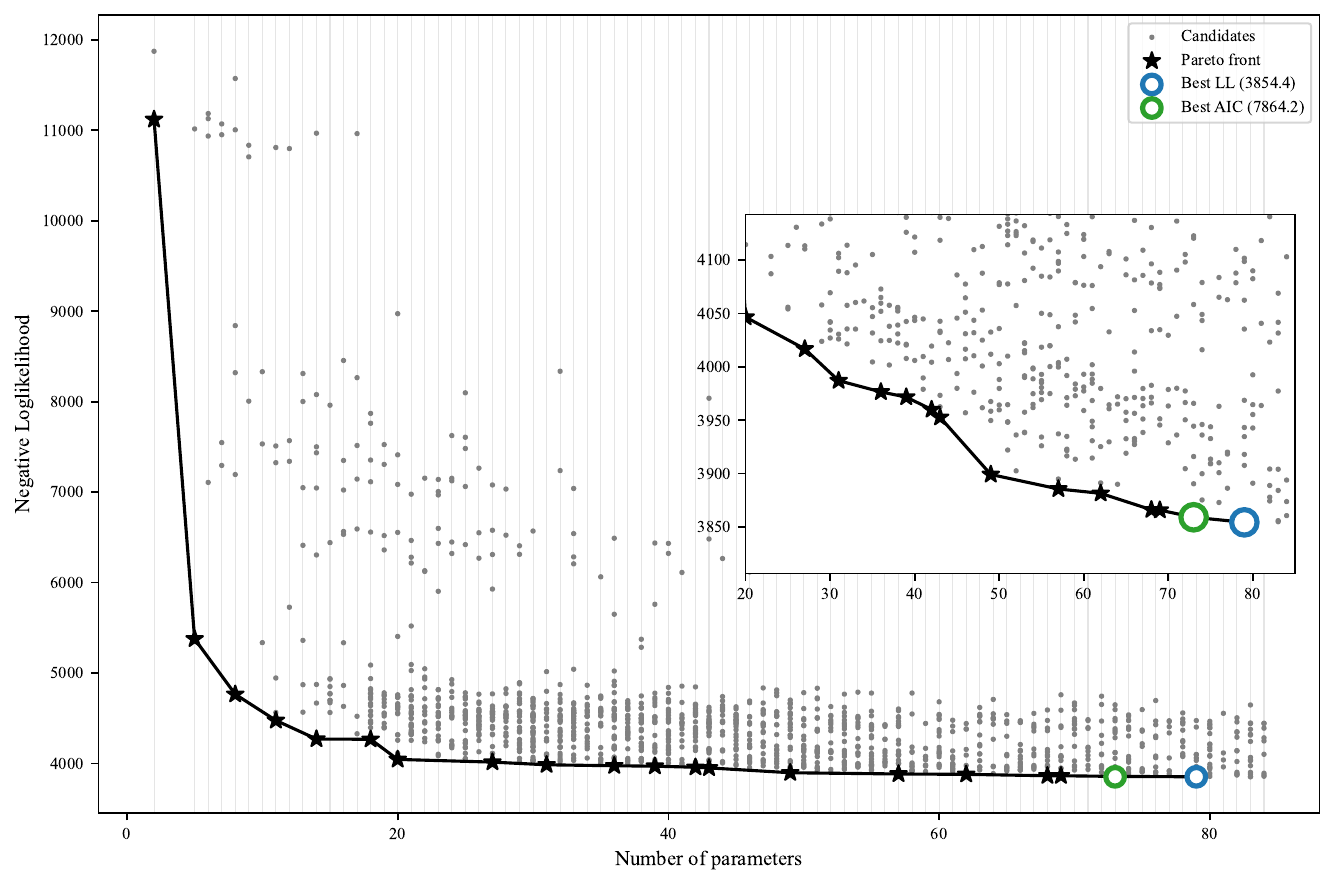}
    \caption{Pareto front of Delphos candidates on the DECISIONS dataset. Grey dots show all converged specifications proposed during training. Stars indicate the non-dominated candidates, and circles show the best AIC and best log-likelihood specifications.}
    \label{fig:exp_decisions_pareto}
\end{figure}

\noindent As in the previous experiment, we built the Pareto front using the unique model specifications suggested by Delphos\footnote{We report the best Pareto front across seeds; the remaining fronts are available in the repository}. Figure~\ref{fig:exp_decisions_pareto} presents the trade-off between the number of parameters and negative log-likelihood. Each grey dot represents a successfully converged model, while the black stars show the non-dominated models forming the Pareto front. The results show a range of models covering this trade-off, from simpler specifications with slightly lower model fit to more complex specifications with improved log-likelihood, all under a reward function that reflects an analyst’s preference for parsimony and behavioural plausibility. \\

\noindent These results highlight not only the agent’s ability to navigate the modelling space and balance multiple modelling outcome objectives, but also its potential to identify models that are consistent with behavioural expectations.  We emphasise that for Delphos (and for SA, and VNS), the construction of the Pareto front, the filtering of behaviourally invalid candidates, and out-of-sample (OOS) validation are performed strictly as post-processing steps after the search process concludes. Because these methods only know the in-sample objective (e.g., AIC or reward) after model estimation, they inevitably explore specifications that violate predefined theoretical expectations. These inconsistent models are subsequently discarded from the search archive a posteriori, while the search dynamics remain unaltered.  Similarly, out-of-sample validation is conducted post-search, evaluating only the subset of candidates that reside on this final Pareto front. In our case, we retain inconsistent model here to provide a complete view of the model fit–complexity trade-off among converged candidates. Ultimately, the modeller retains the responsibility to review the valid Pareto-optimal set and select the final candidate that best balances model performance, out-of-sample generalisation, and theoretical plausibility.

\section{Conclusions}\label{sec: conclusion}

This study introduced Delphos, a reinforcement learning framework for assisting the model specification process of discrete choice models. Rather than treating model specification as a static optimisation problem, we frame it as a sequential decision-making task in which Delphos learns to build utility specifications through a sequence of modelling decisions (e.g., adding or modifying terms). As the contribution of each modelling decision to the final model performance is unknown in advance and intermediate rewards are zero by construction, the methodological value of the Deep Q-Network lies in two mechanisms: (i) it amortises credit assignment by distributing the estimation feedback across the full sequence of modelling actions; and (ii) the Q-network generalises across partial specifications, turning modelling experience into a decision rule. Specifically, over repeated episodes, the agent uses past modelling outcomes to refine its policy and prioritise decisions that lead to high-performing and behaviourally plausible models. To support this learning process, we have defined and developed: (i) a compact yet flexible encoding of the modelling space that captures simple and complex evolving functional forms; (ii) a feasible action space with a masking mechanism; (iii) a user-configurable reward function that can integrate model fit criteria and behavioural expectations; and (iv) an environment that enables sequential interaction between agent actions and model estimation.\\

Delphos not only offers a methodological contribution but also represents a step towards intelligent assistance for discrete choice modelling. While traditional meta-heuristics typically adopt a candidate-centric perspective, applying predefined search operators to propose and evaluate complete specifications without accumulating modelling knowledge across model attempts, Delphos operates at the level of individual modelling decisions and learns from their outcomes. By updating its Q-values from feedback, the agent progressively learns which modelling refinements tend to yield high-performing specifications and uses this accumulated experience to guide subsequent decisions. In this sense, Delphos moves beyond automating the search over model specifications towards learning how to assist the specification process itself.\\

To evaluate its performance, Delphos was applied to both simulated and empirical datasets, including Swissmetro and DECISIONS.  In the simulated cases, our findings demonstrate that the agent can explore large and complex modelling spaces and, even under limited training episodes, suggest well-performing model candidates. The learning curves reveal that the agent rapidly improves its policy in early episodes. As training progresses, the agent increasingly exploits past experience to inform Q-values that shape its search towards high-reward specifications. Moreover, exploration coverage metrics reveal that although the agent does not exhaustively cover the feasible model space, it narrows its set of actions and concentrates its exploration in regions associated with higher rewards. In the empirical cases, we benchmarked Delphos against Variable Neighbourhood Search (VNS) and Simulated Annealing (SA) on the Swissmetro dataset under identical experimental conditions. Our findings showed that Delphos achieved the lowest AIC across seed and reached competitive performance within the first 130 model attempts. While SA dominates the low-complexity range, Delphos extends the frontier towards more complex specifications with lower AIC values. This result is consistent with Delphos' ability to immediately construct complex specifications by accommodating non-linearities and observed heterogeneity in a single episode, without requiring local movements from previously evaluated candidates. Finally, Delphos also proposed specifications that generalise on a held-out set, outperforming the evaluated metaheuristics. On Decisions, we showed that Delphos is capable of having multiple modelling outcomes and incorporating domain-specific behavioural expectations into its learning process. After training, the resulting candidate models can be used to derive Pareto frontiers, offering choice modellers a set of specifications that meet trade-offs between competing goals and supporting informed model selection.\\

Despite these contributions, the proposed framework has several limitations. First, Delphos is currently designed to work on a single-choice dataset, which limits the transferability of modelling strategies across related contexts. As a result, each new dataset requires a new search process, as in existing meta-heuristic approaches. This limitation could be addressed by moving towards a multitask setting, in which the agent learns a reusable policy across multiple choice datasets and adapts it to each specific context through representation learning \citep{rusu2015policy, huang2020closer}. Second, the agent may initially produce behaviourally implausible models, particularly when faced with conflicting objectives in the reward function. Future improvements could involve offline learning methods, where the replay buffer is fed with a known sequence of actions that generate plausible specifications. Third, the agent currently samples past experiences uniformly, which may lead to inefficient learning from low-reward actions. To address this, future implementations could adopt prioritised experience replay \citep{schaul2015prioritized}, allowing the agent to focus on learning from high-value transitions.\\

From an algorithmic standpoint, we implemented this framework using a baseline and single-agent RL algorithm. Future research could benefit from more advanced and robust algorithms that support parallelism and stabilised training. For example, Asynchronous Advantage Actor-Critic (A3C) methods not only allow concurrent learning across multiple agents, thereby improving computational efficiency, but also reduce training instability by incorporating an advanced baseline to regularise Q-value functions \citep{mnih2016asynchronous}. Beyond single-agent architectures, Multi-Agent Reinforcement Learning (MARL) can also be deployed in parallel, each specialised in different modelling outcomes, and collaboratively optimise a shared policy within a common environment \citep{zhang2021multi}. These configurations could enhance exploration, reduce computation time, and enable coordinated learning across diverse modelling goals. \\

Overall, beyond its specific implementation, this work paves the way for a new generation of intelligent tools for discrete choice modelling. Delphos illustrates how reinforcement learning agents can develop adaptive strategies by leveraging feedback and accumulated experience. This complements emerging efforts such as \citet{sfeir2026can}, which explore the use of Large Language Models to suggest and estimate MNL specifications based on domain context and implicit common-sense reasoning. Together, these developments point towards a broader shift from using computational methods merely to execute predefined modelling procedures towards AI systems that can actively assist modellers throughout the modelling process. While this study focuses specifically on utility specification, future intelligent modelling assistants could potentially support other stages of the choice-modelling workflow, including data preparation, experimental design, model estimation and validation. Such systems should not replace the modeller’s theoretical reasoning and domain expertise, but could increasingly take over routine exploration and execution while allowing the modeller to guide, evaluate, and refine the modelling process. Delphos provides an initial step in this direction by demonstrating how modelling experience can be learned and reused to assist discrete choice model specification.

\section*{Acknowledgements}
Gabriel Nova gratefully acknowledges Matthijs Spaan for his early insights and constructive feedback. The first author used ChatGPT for coding and proofreading. All content was reviewed and edited by the authors, who take full responsibility for the publication. Stephane Hess acknowledges the financial support by the European Research Council through the advanced grant 101020940-SYNERGY. 
\clearpage


\appendix
\section{Example of Action Masking in the Specification Process}\label{appendix: masked actions}

The agent uses a masking mechanism that filters invalid actions based on the current specification. At each step, the environment evaluates the existing state $S_e$ and masks out actions that would either violate modelling logic (e.g., duplicating a covariate-transformation pair) or structural constraints (e.g., attempting to modify an unadded variable). Let us consider the following sequence of internal episodes (states) and their corresponding allowed actions:

\begin{align*}
    S^0_e &= [\;] &&\rightarrow\; \text{(add, 1, linear, 0, 0)} \\
    S^1_e &= [(1,\text{linear}, 0, 0)] &&\rightarrow\; \text{(change, 1, linear, 1, 0)} \\
    S^2_e &= [(1,\text{linear}, 1, 0)] &&\rightarrow\; \text{\sout{(add, 1, linear, 0, 0)}}, \text{(add, 3, linear, 0, 0)} \\
    S^3_e &= [(1,\text{linear}, 1, 0), (3,\text{linear}, 0, 0)] &&\rightarrow\; \text{(change, 3, log, 0, 0)} \\
    S^4_e &= [(1,\text{linear}, 1, 0), (3,\text{log}, 0, 0)] &&\rightarrow\; \text{(terminate)}
\end{align*}

\noindent
The interpretation of this progression is as follows:

\begin{itemize}
    \item In $S^0_e$, the model is empty. The valid actions are any of adding a new variable. Here, the agent selects \texttt{(add, 1, linear, 0, 0)}.

    \item In $S^1_e$, variable 1 has been added with a linear transformation, generic and without interacting with a covariate. The agent changes it to alternative-specific taste parameter by choosing \texttt{(change, 1, linear, 1, 0)}.

    \item In $S^2_e$, since variable 1 in linear form is already present, the action \texttt{(add, 1, linear, 0, 0)} is masked out. Instead, the agent selects \texttt{(add, 3, linear, 0, 0)}.

    \item In $S^3_e$, a second variable has been added. The agent now modifies it by applying a log transformation: \texttt{(change, 3, log, 0, 0)}.

    \item In $S^4_e$, the agent ends the model specification process and triggers a \texttt{terminate} action.
\end{itemize}

\clearpage
\section{True parameter values for simulated choice datasets}\label{appendix: True_parameters}

\begin{table}[ht]
\centering
\caption{True parameter values for each simulated case}
\label{tab:true-parameters}
\resizebox{0.95\textwidth}{!}{
\begin{tabular}{c|cc|cc|cc}
\hline
\textbf{Experiment} & \multicolumn{2}{c|}{\textbf{$1A$}} & \multicolumn{2}{c|}{\textbf{$1B$}} & \multicolumn{2}{c|}{\textbf{$1C$}} \\
\hline
ASC$_1$
 & $\beta_{\text{ASC1}}$ & $-0.95$
 & $\beta_{\text{ASC1}}$ & $-0.95$
 & $\beta_{\text{ASC1,cov1}=0}$ & $-0.90$ \\
 & & &
 & & $\beta_{\text{ASC1,cov1}=1}$ & $0.43$ \\
\hline
ASC$_2$
 & $\beta_{\text{ASC2}}$ & $0.84$
 & $\beta_{\text{ASC2}}$ & $0.84$
 & $\beta_{\text{ASC2,cov1}=0}$ & $-0.83$ \\
 & &
 & &
 & $\beta_{\text{ASC2,cov1}=1}$ & $0.43$ \\
\hline
$x_1$
 & $\beta_1$ & $-0.72$
 & $\beta_1$ & $-1.65$
 & $\beta_{1,1}$ & $-0.50$ \\
 & &
 & &
 & $\beta_{1,2}$ & $-0.40$ \\
 & & &
 & & $\beta_{1,3}$ & $-0.60$ \\
\hline
$x_2$
 & $\beta_2$ & $0.83$
 & $\beta_{2,1}$ & $-1.33$
 & $\beta_{2,1,\text{cov2}=1}$ & $-1.40$ \\
 & &
 & $\beta_{2,2}$ & $-1.33$
 & $\beta_{2,1,\text{cov2}=2}$ & $-0.97$ \\
 & &
 & $\beta_{2,3}$ & $-1.33$
 & $\beta_{2,1,\text{cov2}=3}$ & $-0.06$ \\
 & &
 & &
 & $\beta_{2,2,\text{cov2}=1}$ & $-1.16$ \\
 & &
 & &
 & $\beta_{2,2,\text{cov2}=2}$ & $-0.85$ \\
 & &
 & &
 & $\beta_{2,2,\text{cov2}=3}$ & $-0.15$ \\
 & &
 & &
 & $\beta_{2,3,\text{cov2}=1}$ & $-1.15$ \\
 & &
 & &
 & $\beta_{2,3,\text{cov2}=2}$ & $-0.65$ \\
 & &
 & &
 & $\beta_{2,3,\text{cov2}=3}$ & $-0.09$ \\
\hline
$x_3$
 & $\beta_3$ & $1.77$
 & $\beta_3$ & $0.57$
 & $\beta_3$ & $-1.40$ \\
 & $\lambda_3$ &
 & $\lambda_3$ & $0.64$
 & & \\
\hline
$x_4$
 & $\beta_4$ & $1.01$
 & $\beta_{4,1}$ & $0.92$
 & $\beta_4$ & $1.70$ \\
 & $\lambda_4$ & $0.20$
 & $\beta_{4,2}$ & $0.92$
 & $\lambda_4$ & $0.09$ \\
 & &
 & $\beta_{4,3}$ & $0.92$
 & & \\
\hline
$x_5$
 & -- & --
 & $\beta_5$ & $1.90$
 & $\beta_5$ & $1.90$ \\
\hline
$x_6$
 & -- & --
 & -- & --
 & -- & -- \\
\hline
\end{tabular}}
\end{table}

\clearpage
\section{Hyperparameter tuning and learning analysis}\label{appendix: hyperparameter}

Delphos performance is sensitive to its hyperparameters and a well-chosen configuration thus can significantly improve its final performance and speed up its learning, as other RL agents \citep{henderson2018deep}. To investigate the hyperparmeters' influence on the agent's learning and performance, we conduct a random search over both the neural network architecture (e.g., layers and neurons) and learning parameters (e.g. learning rate, buffer size, exploration schedule). To do so, we sample 15 configurations from the predefined hyperparameter space (Table~\ref{tab:hyperparam_ranges}) for each simulated experiment, run each configuration using five random seeds\footnote{For all remaining hyperparameters, we use the values used in the original DQN paper \citep{mnih2015human}}, and evaluate them using the area under the learning curve (AUC) and the mean reward in the last 100 episodes (Final--100), following \citet{teh2017distral} to capture both sample-efficiency and asymptotic performance.\\

\begin{table}[h!]
    \centering
    \caption{Hyperparameter ranges.}
    \label{tab:hyperparam_ranges}
    \begin{tabular}{ll}
        \hline
        \textbf{Hyperparameter}          & \textbf{Level values} \\
        \hline
        Number of hidden layers $(\mathbf{L})$           & $\{2,\,3\}$ \\
        Hidden units per layer $(\mathbf{U})$            & $\{64,\,128,\,256,\,512\}$ \\
        Learning rate $(\boldsymbol{\alpha})$                & $\{1e-3, \, 5e-4,\, 1e-4\}$\\
        Discount factor $(\boldsymbol{\gamma})$              & $\{0.95,\,0.99\}$\\
        Replay buffer size $(\mathbf{BS})$               & $\{5{,}000,\,10{,}000,\,20{,}000\}$ \\
        Target network update frequency $(\mathbf{TU})$  & $\{5,\,10,\}$ \\
        Total episodes $(\mathbf{E})$                    & $\{10{,}000,\,30{,}000,\,50{,}000\}$ \\
        \hline
    \end{tabular}
\end{table}

Table~\ref{tab:hp_performance} shows the AUC and Final--100 of the five best-performing sampled hyperparameter configuration for each simulated experiment ($1A$--$1C$). For each configuration, we show the mean, the 25th, and 75th percentiles of the learning curves from the five random seeds.

\begin{table}[ht]
\centering
\scriptsize
\caption{Performance of sampled hyperparameter configurations.}
\label{tab:hp_performance}
\begin{tabular}{lccccccccc}
\toprule
\textbf{Exp}  & \textbf{L} & \textbf{U} & $\boldsymbol{\alpha}$ & $\boldsymbol{\gamma}$ & 
\textbf{BS} & \textbf{TU} & \textbf{E} &  \textbf{AUC} & \textbf{Final--100}  \\
\midrule

$1A$ - conf. 1   & 2 & 256 & 1e--4 & 0.99 & 5k  & 10 & 50k & 0.552 & 0.612 \\
$1A$ - conf. 2   & 3 & 256 & 1e--4 & 0.99 & 5k  & 5  & 50k & 0.553 & 0.612 \\
$1A$ - conf. 3   & 2 & 64  & 1e--4 & 0.99 & 10k & 10 & 5k  & 0.538 & 0.611 \\
\textbf{S\(_1\) - conf. 4}   & \textbf{2} & \textbf{64}  & \textbf{1e--4} & \textbf{0.99} & \textbf{10k} & \textbf{10} & \textbf{10k} & \textbf{0.548} & \textbf{0.611} \\
$1A$ - conf. 5   & 2 & 256 & 1e--4 & 0.95 & 10k & 5  & 30k & 0.538 & 0.610 \\\hline

$1B$ - conf. 1   & 2 & 128 & 1e--4 & 0.99 & 10k & 10  & 50k & 0.708 & 0.753 \\
$1B$ - conf. 2   & 2 & 128 & 1e--4 & 0.99 & 5k & 10 & 50k & 0.707 & 0.753 \\
\textbf{S\(_2\) - conf. 3}   & \textbf{2} & \textbf{128} & \textbf{1e--4} & \textbf{0.99} & \textbf{5k}  & \textbf{10}  & \textbf{10k} & \textbf{0.698} & \textbf{0.753} \\
$1B$ - conf. 4   & 2 & 128 & 1e--4 & 0.95 & 50k & 5 & 10k & 0.686 & 0.719 \\
$1B$ - conf. 5   & 3 & 128 & 5e--4 & 0.95 & 5k  & 10 & 30k & 0.684 & 0.722 \\\hline

\textbf{S\(_3\) - conf. 1}   & \textbf{2} & \textbf{256} & \textbf{1e--4} & \textbf{0.99} & \textbf{5k}    & \textbf{5}     & \textbf{10k} & \textbf{0.516} & \textbf{0.657} \\
$1C$ - conf. 2   & 2 & 256 & 1e--3 & 0.99 & 5k    & 10    & 50k & 0.503 & 0.653 \\
$1C$ - conf. 3   & 3 & 128 & 1e--4 & 0.99 & 5k    & 10    & 10k & 0.511 & 0.648 \\
$1C$ - conf. 4   & 2 & 128 & 5e--3 & 0.99 & 5k    & 5     & 10k & 0.474 & 0.644 \\
$1C$ - conf. 5   & 3 & 128 & 1e--4 & 0.99 & 10k   & 5     & 10k & 0.476 & 0.643 \\

\bottomrule
\end{tabular}
\end{table}

For our results reported in Section~\ref{sec:simulated_experiments}, we favour configurations that achieve high performance while requiring fewer training episodes. Within the top five candidates for each experiment, we thus select the configuration that offers a favourable trade-off between sample efficiency and asymptotic performance. For Experiments~$1A, 1B$, and $1C$, we respectively select \emph{$1A$--conf.~4}, \emph{$1B$--conf.~3}, and \emph{$1C$--conf.~1}, which reduce the training horizon to $E = 10{,}000$ episodes while still obtaining competitive Final--100 rewards (within 0.02 of the best configuration in each case).

\clearpage
\section{Simulated experiment $1B$ and $1C$}\label{appendix: simulated experiment 1B and 1C}

\begin{figure}[ht]
    \centering
    \includegraphics[width=\linewidth]{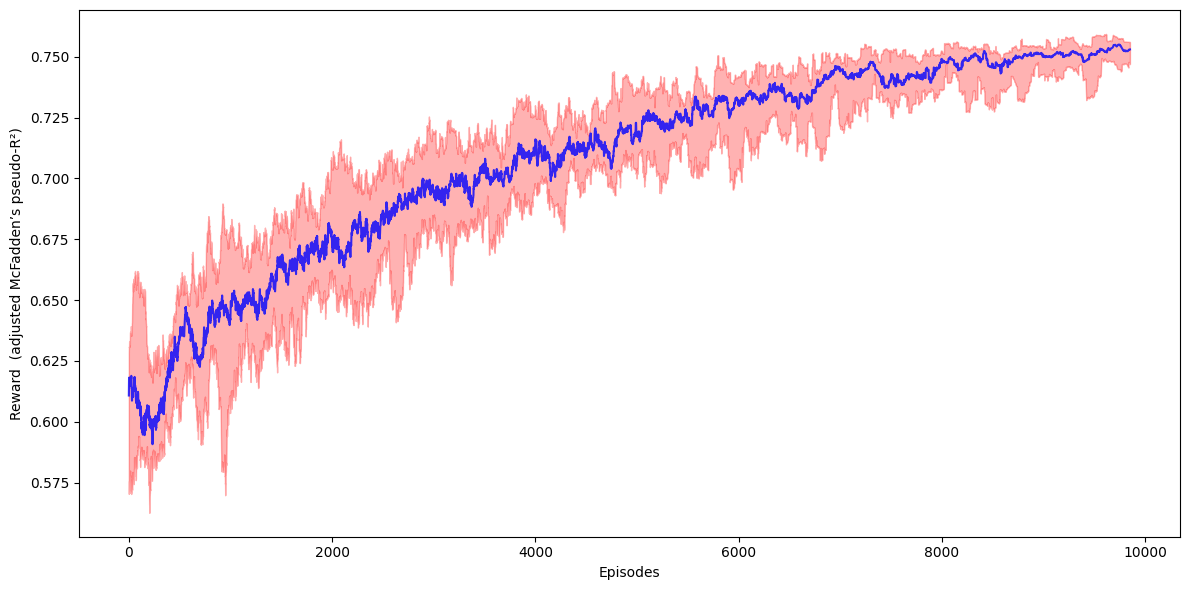}
    \caption{Learning curve showing the reward signal ($\bar{\rho}^2$) across 10 runs}
    \label{fig: learning_curve_1b}
\end{figure}

\begin{figure}[ht]
    \centering
    \includegraphics[width=\linewidth]{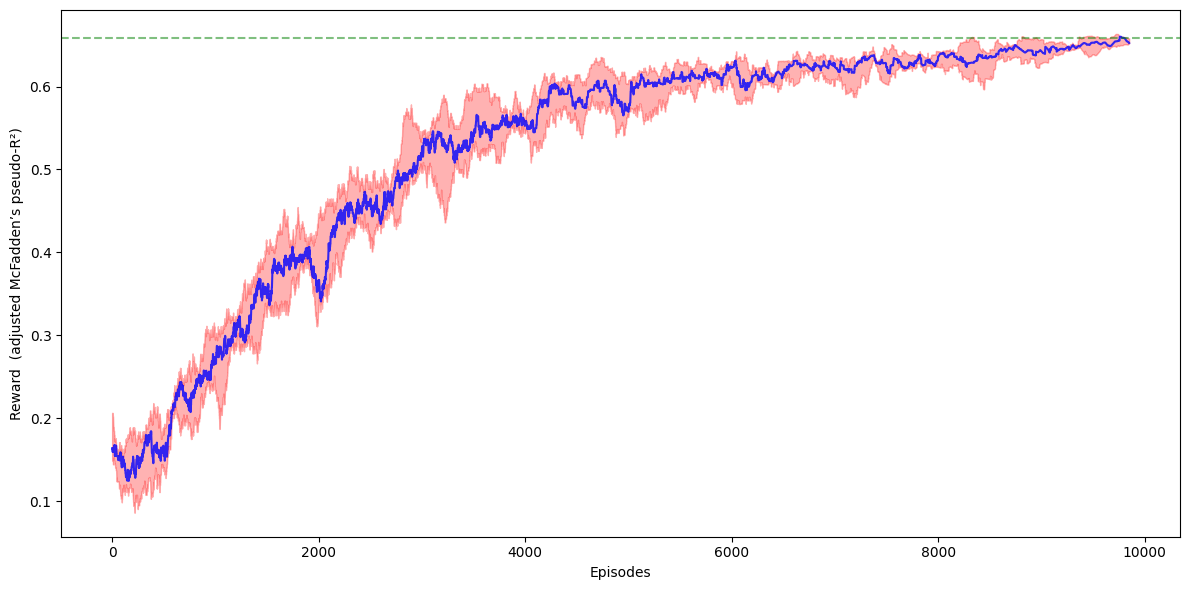}
    \caption{Learning curve showing the reward signal ($\bar{\rho}^2$) across 10 runs}
    \label{fig: learning_curve_1c}
\end{figure}
\clearpage
\section{Simulated experiment A: action space} \label{appendix:action-space}

\begin{table}[ht]
\centering
\caption{Action space for simulated experiment $1A$}
\label{tab:action_space}
\begin{tabular}{clcll}
\toprule
\textbf{Action ID} & \textbf{Operation} & \textbf{Variable} & \textbf{Transformation Type} & \textbf{Action} \\
\midrule
0  & terminate & 0 & -       & (\texttt{end}) \\
1  & add       & 1 & linear   & (\texttt{add}, 1, \texttt{linear}) \\
2  & change    & 1 & linear   & (\texttt{change}, 1, \texttt{linear}) \\
3  & add       & 1 & log      & (\texttt{add}, 1, \texttt{log}) \\
4  & change    & 1 & log      & (\texttt{change}, 1, \texttt{log}) \\
5  & add       & 1 & box-cox  & (\texttt{add}, 1, \texttt{box-cox}) \\
6  & change    & 1 & box-cox  & (\texttt{change}, 1, \texttt{box-cox}) \\
7  & add       & 2 & linear   & (\texttt{add}, 2, \texttt{linear}) \\
8  & change    & 2 & linear   & (\texttt{change}, 2, \texttt{linear}) \\
9  & add       & 2 & log      & (\texttt{add}, 2, \texttt{log}) \\
10 & change    & 2 & log      & (\texttt{change}, 2, \texttt{log}) \\
11 & add       & 2 & box-cox  & (\texttt{add}, 2, \texttt{box-cox}) \\
12 & change    & 2 & box-cox  & (\texttt{change}, 2, \texttt{box-cox}) \\
13 & add       & 3 & linear   & (\texttt{add}, 3, \texttt{linear}) \\
14 & change    & 3 & linear   & (\texttt{change}, 3, \texttt{linear}) \\
15 & add       & 3 & log      & (\texttt{add}, 3, \texttt{log}) \\
16 & change    & 3 & log      & (\texttt{change}, 3, \texttt{log}) \\
17 & add       & 3 & box-cox  & (\texttt{add}, 3, \texttt{box-cox}) \\
18 & change    & 3 & box-cox  & (\texttt{change}, 3, \texttt{box-cox}) \\
19 & add       & 4 & linear   & (\texttt{add}, 4, \texttt{linear}) \\
20 & change    & 4 & linear   & (\texttt{change}, 4, \texttt{linear}) \\
21 & add       & 4 & log      & (\texttt{add}, 4, \texttt{log}) \\
22 & change    & 4 & log      & (\texttt{change}, 4, \texttt{log}) \\
23 & add       & 4 & box-cox  & (\texttt{add}, 4, \texttt{box-cox}) \\
24 & change    & 4 & box-cox  & (\texttt{change}, 4, \texttt{box-cox}) \\
25 & add       & 5 & linear   & (\texttt{add}, 5, \texttt{linear}) \\
26 & change    & 5 & linear   & (\texttt{change}, 5, \texttt{linear}) \\
27 & add       & 5 & log      & (\texttt{add}, 5, \texttt{log}) \\
28 & change    & 5 & log      & (\texttt{change}, 5, \texttt{log}) \\
29 & add       & 5 & box-cox  & (\texttt{add}, 5, \texttt{box-cox}) \\
30 & change    & 5 & box-cox  & (\texttt{change}, 5, \texttt{box-cox}) \\
31 & add       & 6 & linear   & (\texttt{add}, 6, \texttt{linear}) \\
32 & change    & 6 & linear   & (\texttt{change}, 6, \texttt{linear}) \\
33 & add       & 6 & log      & (\texttt{add}, 6, \texttt{log}) \\
34 & change    & 6 & log      & (\texttt{change}, 6, \texttt{log}) \\
35 & add       & 6 & box-cox  & (\texttt{add}, 6, \texttt{box-cox}) \\
36 & change    & 6 & box-cox  & (\texttt{change}, 6, \texttt{box-cox}) \\
\bottomrule
\end{tabular}
\end{table}

\clearpage
\section{Swissmetro: Modelling outcomes}\label{appendix: modelling results Delphos best AIC}

\begin{longtable}{lccr}
\caption{Best-AIC specification on Swissmetro dataset}
\label{tab:estimates_swissmetro_delphos}\\
\hline
Parameter & Transformation & Covariate & Value (Rob. s.e.) \\
\hline
$\beta_{\text{train},\,\text{age}=1}$ & & age = 1 & $1.483\,(0.706)$ \\
$\beta_{\text{train},\,\text{age}=2}$ & & age = 2 & $0.515\,(0.371)$ \\
$\beta_{\text{train},\,\text{age}=3}$ & & age = 3 & $0.232\,(0.399)$ \\
$\beta_{\text{train},\,\text{age}=4}$ & & age = 4 & $0.746\,(0.482)$ \\
$\beta_{\text{train},\,\text{age}=5}$ & & age = 5 & $2.168\,(0.499)$ \\
$\beta_{\text{sm},\,\text{age}=1}$ & & age = 1 & $0.682\,(0.879)$ \\
$\beta_{\text{sm},\,\text{age}=2}$ & & age = 2 & $0.315\,(0.514)$ \\
$\beta_{\text{sm},\,\text{age}=3}$ & & age = 3 & $0.245\,(0.515)$ \\
$\beta_{\text{sm},\,\text{age}=4}$ & & age = 4 & $-0.043\,(0.553)$ \\
$\beta_{\text{sm},\,\text{age}=5}$ & & age = 5 & $-0.012\,(0.616)$ \\
$\beta_{\text{train},TT,\,\text{class}=0}$ & Box--Cox & class = 0 & $-1.114\,(0.367)$ \\
$\beta_{\text{train},TT,\,\text{class}=1}$ & Box--Cox & class = 1 & $-1.875\,(0.408)$ \\
$\beta_{\text{sm},TT,\,\text{class}=0}$ & Box--Cox & class = 0 & $-1.203\,(0.314)$ \\
$\beta_{\text{sm},TT,\,\text{class}=1}$ & Box--Cox & class = 1 & $-1.540\,(0.236)$ \\
$\beta_{\text{car},TT,\,\text{class}=0}$ & Box--Cox & class = 0 & $-1.377\,(0.326)$ \\
$\beta_{\text{car},TT,\,\text{class}=1}$ & Box--Cox & class = 1 & $-2.484\,(0.310)$ \\
$\lambda_{TT}$ & & & $0.101\,(0.228)$ \\
$\beta_{\text{train},TC,\,\text{gender}=0}$ & Log & gender = 0 & $-3.066\,(1.201)$ \\
$\beta_{\text{train},TC,\,\text{gender}=1}$ & Log & gender = 1 & $-6.638\,(0.831)$ \\
$\beta_{\text{sm},TC,\,\text{gender}=0}$ & Log & gender = 0 & $-2.070\,(0.961)$ \\
$\beta_{\text{sm},TC,\,\text{gender}=1}$ & Log & gender = 1 & $-3.483\,(0.478)$ \\
$\beta_{\text{car},TC,\,\text{gender}=0}$ & Log & gender = 0 & $-1.207\,(1.220)$ \\
$\beta_{\text{car},TC,\,\text{gender}=1}$ & Log & gender = 1 & $-2.232\,(0.504)$ \\
$\beta_{\text{train},HE,\,\text{income}=1}$ & Box--Cox & income = 1 & $-0.325\,(0.228)$ \\
$\beta_{\text{train},HE,\,\text{income}=2}$ & Box--Cox & income = 2 & $-0.827\,(0.278)$ \\
$\beta_{\text{train},HE,\,\text{income}=3}$ & Box--Cox & income = 3 & $-0.411\,(0.337)$ \\
$\beta_{\text{train},HE,\,\text{income}=4}$ & Box--Cox & income = 4 & $-1.267\,(0.405)$ \\
$\beta_{\text{sm},HE,\,\text{income}=1}$ & Box--Cox & income = 1 & $-0.271\,(0.481)$ \\
$\beta_{\text{sm},HE,\,\text{income}=2}$ & Box--Cox & income = 2 & $-0.569\,(0.576)$ \\
$\beta_{\text{sm},HE,\,\text{income}=3}$ & Box--Cox & income = 3 & $-0.774\,(0.708)$ \\
$\beta_{\text{sm},HE,\,\text{income}=4}$ & Box--Cox & income = 4 & $-0.472\,(0.576)$ \\
$\lambda_{HE}$ & & & $0.770\,(0.525)$ \\
$\beta_{SE,\,\text{ga}=0}$ & Log & GA = 0 & $0.507\,(0.270)$ \\
$\beta_{SE,\,\text{ga}=1}$ & Log & GA = 1 & $-0.782\,(0.353)$ \\ \hline
LL($0$) & \multicolumn{3}{r}{-5,548.280} \\
LL(final) & \multicolumn{3}{r}{-3,692.74} \\
AIC & \multicolumn{3}{r}{7,453.47} \\
BIC & \multicolumn{3}{r}{7,677.73} \\
Rho-squared & \multicolumn{3}{r}{0.334} \\
Adj.\ Rho-squared & \multicolumn{3}{r}{0.328} \\
N.\ observations & \multicolumn{3}{r}{5,409} \\
Number of parameters & \multicolumn{3}{r}{34} \\\hline
\end{longtable}

\clearpage

\section{Decisions: Modelling outcomes}\label{appendix: best_LL_params_decisions}
\begin{longtable}{lccr}
\caption{Best-LL\_Nparams specification on Decisions dataset.\label{tab:estimates_decisions_delphos}}
\label{tab:estimates_model_2}\\
\hline
Parameter & Transformation & Covariate & Value (Rob. s.e.) \\
\hline

$\beta_{\text{car},\,\text{edu}=1}$ & & edu = 1 & $-0.521\,(0.394)$ \\
$\beta_{\text{car},\,\text{edu}=2}$ & & edu = 2 & $-0.594\,(0.348)$ \\
$\beta_{\text{car},\,\text{edu}=3}$ & & edu = 3 & $-0.887\,(0.248)$ \\
$\beta_{\text{car},\,\text{edu}=4}$ & & edu = 4 & $-1.654\,(0.201)$ \\
$\beta_{\text{car},\,\text{edu}=5}$ & & edu = 5 & $-1.661\,(0.237)$ \\
$\beta_{\text{car},\,\text{edu}=6}$ & & edu = 6 & $-2.038\,(0.275)$ \\
$\beta_{\text{bus},\,\text{edu}=1}$ & & edu = 1 & $-0.035\,(0.305)$ \\
$\beta_{\text{bus},\,\text{edu}=2}$ & & edu = 2 & $0.126\,(0.260)$ \\
$\beta_{\text{bus},\,\text{edu}=3}$ & & edu = 3 & $-0.345\,(0.237)$ \\
$\beta_{\text{bus},\,\text{edu}=4}$ & & edu = 4 & $-1.149\,(0.207)$ \\
$\beta_{\text{bus},\,\text{edu}=5}$ & & edu = 5 & $-1.132\,(0.224)$ \\
$\beta_{\text{bus},\,\text{edu}=6}$ & & edu = 6 & $-1.231\,(0.270)$ \\
$\beta_{\text{rail},\,\text{edu}=1}$ & & edu = 1 & $2.388\,(0.740)$ \\
$\beta_{\text{rail},\,\text{edu}=2}$ & & edu = 2 & $3.658\,(0.666)$ \\
$\beta_{\text{rail},\,\text{edu}=3}$ & & edu = 3 & $2.702\,(0.594)$ \\
$\beta_{\text{rail},\,\text{edu}=4}$ & & edu = 4 & $2.235\,(0.570)$ \\
$\beta_{\text{rail},\,\text{edu}=5}$ & & edu = 5 & $2.709\,(0.592)$ \\
$\beta_{\text{rail},\,\text{edu}=6}$ & & edu = 6 & $1.468\,(0.682)$ \\
$\beta_{\text{taxi},\,\text{edu}=1}$ & & edu = 1 & $-2.042\,(0.606)$ \\
$\beta_{\text{taxi},\,\text{edu}=2}$ & & edu = 2 & $-3.510\,(0.646)$ \\
$\beta_{\text{taxi},\,\text{edu}=3}$ & & edu = 3 & $-2.788\,(0.557)$ \\
$\beta_{\text{taxi},\,\text{edu}=4}$ & & edu = 4 & $-2.972\,(0.523)$ \\
$\beta_{\text{taxi},\,\text{edu}=5}$ & & edu = 5 & $-3.971\,(0.564)$ \\
$\beta_{\text{taxi},\,\text{edu}=6}$ & & edu = 6 & $-3.261\,(0.560)$ \\
$\beta_{\text{cycling},\,\text{edu}=1}$ & & edu = 1 & $-4.111\,(0.427)$ \\
$\beta_{\text{cycling},\,\text{edu}=2}$ & & edu = 2 & $-4.506\,(0.371)$ \\
$\beta_{\text{cycling},\,\text{edu}=3}$ & & edu = 3 & $-4.062\,(0.267)$ \\
$\beta_{\text{cycling},\,\text{edu}=4}$ & & edu = 4 & $-4.712\,(0.201)$ \\
$\beta_{\text{cycling},\,\text{edu}=5}$ & & edu = 5 & $-5.532\,(0.258)$ \\
$\beta_{\text{cycling},\,\text{edu}=6}$ & & edu = 6 & $-5.844\,(0.360)$ \\
$\beta_{\text{car},IVT,\,\text{hh size}=1}$ & Linear & hh size = 1 & $-0.085\,(0.009)$ \\
$\beta_{\text{car},IVT,\,\text{hh size}=2}$ & Linear & hh size = 2 & $-0.146\,(0.014)$ \\
$\beta_{\text{car},IVT,\,\text{hh size}=3}$ & Linear & hh size = 3 & $-0.118\,(0.020)$ \\
$\beta_{\text{car},IVT,\,\text{hh size}=4}$ & Linear & hh size = 4 & $-0.144\,(0.047)$ \\
$\beta_{\text{car},IVT,\,\text{hh size}=5}$ & Linear & hh size = 5 & $-0.190\,(0.077)$ \\
$\beta_{\text{bus},IVT,\,\text{hh size}=1}$ & Linear & hh size = 1 & $-0.031\,(0.005)$ \\
$\beta_{\text{bus},IVT,\,\text{hh size}=2}$ & Linear & hh size = 2 & $-0.065\,(0.009)$ \\
$\beta_{\text{bus},IVT,\,\text{hh size}=3}$ & Linear & hh size = 3 & $-0.073\,(0.014)$ \\
$\beta_{\text{bus},IVT,\,\text{hh size}=4}$ & Linear & hh size = 4 & $-0.096\,(0.043)$ \\
$\beta_{\text{bus},IVT,\,\text{hh size}=5}$ & Linear & hh size = 5 & $-0.083\,(0.050)$ \\
$\beta_{\text{rail},IVT,\,\text{hh size}=1}$ & Linear & hh size = 1 & $-0.033\,(0.009)$ \\
$\beta_{\text{rail},IVT,\,\text{hh size}=2}$ & Linear & hh size = 2 & $-0.093\,(0.016)$ \\
$\beta_{\text{rail},IVT,\,\text{hh size}=3}$ & Linear & hh size = 3 & $-0.070\,(0.025)$ \\
$\beta_{\text{rail},IVT,\,\text{hh size}=4}$ & Linear & hh size = 4 & $-0.147\,(0.066)$ \\
$\beta_{\text{rail},IVT,\,\text{hh size}=5}$ & Linear & hh size = 5 & $-0.278\,(0.137)$ \\
$\beta_{\text{taxi},IVT,\,\text{hh size}=1}$ & Linear & hh size = 1 & $-0.084\,(0.025)$ \\
$\beta_{\text{taxi},IVT,\,\text{hh size}=2}$ & Linear & hh size = 2 & $-0.151\,(0.030)$ \\
$\beta_{\text{taxi},IVT,\,\text{hh size}=3}$ & Linear & hh size = 3 & $-0.109\,(0.033)$ \\
$\beta_{\text{taxi},IVT,\,\text{hh size}=4}$ & Linear & hh size = 4 & $-0.142\,(0.062)$ \\
$\beta_{\text{taxi},IVT,\,\text{hh size}=5}$ & Linear & hh size = 5 & $-0.111\,(0.085)$ \\
$\beta_{\text{cycling},IVT,\,\text{hh size}=1}$ & Linear & hh size = 1 & $-0.077\,(0.007)$ \\
$\beta_{\text{cycling},IVT,\,\text{hh size}=2}$ & Linear & hh size = 2 & $-0.094\,(0.010)$ \\
$\beta_{\text{cycling},IVT,\,\text{hh size}=3}$ & Linear & hh size = 3 & $-0.083\,(0.014)$ \\
$\beta_{\text{cycling},IVT,\,\text{hh size}=4}$ & Linear & hh size = 4 & $-0.059\,(0.027)$ \\
$\beta_{\text{cycling},IVT,\,\text{hh size}=5}$ & Linear & hh size = 5 & $-0.073\,(0.047)$ \\
$\beta_{\text{walking},IVT,\,\text{hh size}=1}$ & Linear & hh size = 1 & $-0.155\,(0.006)$ \\
$\beta_{\text{walking},IVT,\,\text{hh size}=2}$ & Linear & hh size = 2 & $-0.155\,(0.008)$ \\
$\beta_{\text{walking},IVT,\,\text{hh size}=3}$ & Linear & hh size = 3 & $-0.164\,(0.010)$ \\
$\beta_{\text{walking},IVT,\,\text{hh size}=4}$ & Linear & hh size = 4 & $-0.162\,(0.018)$ \\
$\beta_{\text{walking},IVT,\,\text{hh size}=5}$ & Linear & hh size = 5 & $-0.128\,(0.022)$ \\
$\beta_{\text{bus},OVT,\,\text{city}=1}$ & Log & city = 1 & $-1.518\,(0.083)$ \\
$\beta_{\text{bus},OVT,\,\text{city}=2}$ & Log & city = 2 & $-1.427\,(0.099)$ \\
$\beta_{\text{bus},OVT,\,\text{city}=3}$ & Log & city = 3 & $-1.934\,(0.110)$ \\
$\beta_{\text{rail},OVT,\,\text{city}=1}$ & Log & city = 1 & $-2.754\,(0.188)$ \\
$\beta_{\text{rail},OVT,\,\text{city}=2}$ & Log & city = 2 & $-1.657\,(0.206)$ \\
$\beta_{\text{rail},OVT,\,\text{city}=3}$ & Log & city = 3 & $-2.223\,(0.203)$ \\
$\beta_{\text{car},TC,\,\text{city}=1}$ & Box--Cox & city = 1 & $-0.943\,(0.087)$ \\
$\beta_{\text{car},TC,\,\text{city}=2}$ & Box--Cox & city = 2 & $-0.773\,(0.172)$ \\
$\beta_{\text{car},TC,\,\text{city}=3}$ & Box--Cox & city = 3 & $-0.838\,(0.118)$ \\
$\beta_{\text{bus},TC,\,\text{city}=1}$ & Box--Cox & city = 1 & $-1.089\,(0.066)$ \\
$\beta_{\text{bus},TC,\,\text{city}=2}$ & Box--Cox & city = 2 & $-0.723\,(0.132)$ \\
$\beta_{\text{bus},TC,\,\text{city}=3}$ & Box--Cox & city = 3 & $-1.077\,(0.174)$ \\
$\beta_{\text{rail},TC,\,\text{city}=1}$ & Box--Cox & city = 1 & $-0.118\,(0.085)$ \\
$\beta_{\text{rail},TC,\,\text{city}=2}$ & Box--Cox & city = 2 & $-1.096\,(0.220)$ \\
$\beta_{\text{rail},TC,\,\text{city}=3}$ & Box--Cox & city = 3 & $-0.679\,(0.153)$ \\
$\beta_{\text{taxi},TC,\,\text{city}=1}$ & Box--Cox & city = 1 & $-1.107\,(0.372)$ \\
$\beta_{\text{taxi},TC,\,\text{city}=2}$ & Box--Cox & city = 2 & $-1.245\,(0.386)$ \\
$\beta_{\text{taxi},TC,\,\text{city}=3}$ & Box--Cox & city = 3 & $-1.224\,(0.380)$ \\
$\lambda_3$ & & & $0.246\,(0.075)$ \\
\hline
LL($0$)                 & \multicolumn{3}{r}{-11,965.38} \\
LL(final)               & \multicolumn{3}{r}{-3,854.37} \\
AIC                     & \multicolumn{3}{r}{7,866.76} \\
BIC                     & \multicolumn{3}{r}{8,436.51} \\
Rho-squared             & \multicolumn{3}{r}{0.678} \\
Adj.\ Rho-squared       & \multicolumn{3}{r}{0.671} \\
N.\ observations        & \multicolumn{3}{r}{10,019} \\
Number of parameters    & \multicolumn{3}{r}{79} \\
\hline
\end{longtable}

\end{document}